\documentclass[twocolumn,amsmath,amssymb,aps, prb,
showkeys,showpacs,nofootinbib] {revtex4-1}
\usepackage{graphicx}
\usepackage{bm}
\usepackage[colorlinks=true,allcolors=blue]{hyperref}
\usepackage{dcolumn}

\usepackage{float}

\begin{document}
\title{Charge-response of the Majorana toric code}
\author{Ananda Roy}
\affiliation{JARA Institute for Quantum Information, RWTH Aachen University, 52056 Aachen, Germany}

\author{Fabian Hassler}
\affiliation{JARA Institute for Quantum Information, RWTH Aachen University, 52056 Aachen, Germany}
\begin{abstract}
At zero temperature, a two dimensional lattice of Majorana zero modes on mesoscopic superconducting islands has a topologically ordered toric code phase. Recently, a Landau field theory has been proposed for the system that captures its different phases and the associated phase-transitions. 
It was shown that with the increase of Josephson tunneling between the islands, a continuous symmetry-breaking 3D-XY transition gets transformed into a discrete symmetry-breaking 3D-Ising transition through a couple of tricritical points and first order transitions.
Using the proposed field theory, we analyze the charge-response of the system at the different continuous phase-transitions. We calculate the universal conductivity at the 3D-XY transitions and the change in the superconducting density at the Ising transition using 1/N expansion. Furthermore, by computing a one-loop correction to the field theory, we show that an additional tricritical point is likely to be present in the phase-diagram. Finally, we provide a mean-field calculation that supports the earlier proposed field theory.
\end{abstract}
\maketitle 

\section{Introduction}
One of the most promising platforms for fault-tolerant quantum computation\cite{DiVincenzo2009} is the toric code proposed by Kitaev.\cite{Kitaev2003, Kitaev2006,Fowler_Cleland_2012} The ground space of the toric code is four-fold degenerate where two qubits can be encoded. The primary advantage of this way of encoding is that the degeneracy of the ground space depends only on the topology of the space on which the toric code is implemented and thus, is robust towards local perturbations. A direct realization of the toric code is by placing qubits on the links of a large square lattice with periodic boundary conditions and performing measurements that project the system to the topologically ordered ground space. \cite{Kitaev2003} Alternately, a Hamiltonian can be designed whose the low-energy sector can be described by the toric code Hamiltonian.\cite{Levin2006, Kitaev2006} This is the case for a system of spins located at the edges of a 2D honeycomb lattice, interacting via alternating $\sigma_x\sigma_x$, $\sigma_y\sigma_y$ and $\sigma_z\sigma_z$ interactions around each plaquette of the lattice. For a suitable choice of interaction strengths, the system is in the toric code phase. \cite{Kitaev2006} 

In contrast to the aforementioned schemes of realizing the toric code using spins/qubits, alternative approaches have been
proposed using Majorana zero modes.\cite{Xu2010, Terhal2012, Nussinov2012, 
Vijay2015, Landau2016, Karzig2016, Litinski2017}  In these approaches, a 2D lattice of mesoscopic superconducting islands is considered. On each of these islands, two Kitaev wires\cite{Kitaev2000} are embedded which contain localized Majorana zero modes. Due to the finite size, each island has a finite charging energy, denoted by $E_C = e^2/2C$, where $C$ is the capacitance of each island to a ground plane. Due to the Josephson effect, the Cooper-pairs tunnel between different neighboring islands at a rate $E_J$. Furthermore, the Majorana zero modes enable tunneling of single electrons between two neighboring islands at a rate $E_M$.\cite{Fu2010} For vanishing Josephson tunneling rate and $E_M\ll E_C$, the system is in the topologically ordered toric code phase.\cite{Xu2010, Landau2016} Upon increasing the single-electron tunneling rate, the system undergoes a topological phase-transition of 3D-XY type into a topologically trivial state.\cite{Xu2010} Moreover, in the limit of infinite Cooper-pair tunneling rate, the system is also in the toric code phase. In this case, the phase-transition to the topologically trivial state is of 3D-Ising type. A Landau field theory analysis for the system has been done in Ref. \onlinecite{Roy2017}. It was shown that upon increasing the Josephson coupling strength from zero, the line of the 3D-XY topological phase-transition terminated at a 3D-XY tricritical point. Subsequently, it turned first order, which then terminated at a 3D-Ising tricritical point. Further increase of the Josephson coupling made the phase-transition a 3D-Ising one. The charge signatures of the different phases are as follows. For $E_J\ll E_M\ll E_C$, the system is a Mott-insulator. Upon Increasing $E_J$, the system undergoes an additional 3D-XY phase-transition into a charge-$2e$ superconductor. \cite{Fisher1989, Fazio1997, Fazio2001, Herbut2007, Sachdev2011} Most importantly, the system stays in the toric code phase both as a Mott-insulator and as a charge-$2e$ superconductor. Upon increasing $E_M$ from either one of these phases, the system makes a topological phase-transition to a charge-$e$ superconductor\cite{Wen1991,Balents1999, Senthil2000}. Depending on the strength of $E_J$, it is the nature of this phase-transition that changes from 3D-XY to 3D-Ising. 

Given that the phases have distinct charge signatures, we investigate the
charge response across the different phase transitions.  Measurement of the
superconducting densities and correlation lengths reveals the critical
exponents.\cite{Roy2017} Most importantly, a measurement of the electrical
conductivity provides a unique signature that distinguishes between the
different phase-transitions. The main goal of this work is to provide
quantitative predictions of the conductivity that are amenable to experimental verification.

In the context of the Bose-Hubbard model at zero temperature, the conductivity at a 3D-XY transition has been shown to be a universal value. This conductivity value has been computed numerically using Monte-Carlo methods\cite{Cha1991} and analytically using 1/N expansion \cite{Cha1991} and $\epsilon$ expansion \cite{Fazio1996}. These results can be directly applied to the two 3D-XY transitions and the 3D-XY tricritical point of our model. However, for the Ising transition and the Ising tricritical point, a fresh computation is necessary. This computation is nontrivial due to the fact that the current is carried by two coupled Ising degrees of freedom, out of which only one undergoes the phase transition. We do this computation to leading order in the interaction using 1/N expansion. We show that as the system undergoes the transition from a charge-$2e$ to a charge-$e$ superconductor, there is a jump in the superconducting density. In contrast to the 3D-XY transitions, there is no dissipative component to the conductivity. 

In addition, we compute a one-loop correction to the field theory describing the system's transition from a Mott-insulator to a charge-$2e$ superconductor. This calculation indicates that another tricritical point of a 3D-XY type is likely to be present in the phase diagram. Lastly, we also perform a mean-field analysis of the model that provides additional support to the Landau field theory proposed in Ref. \onlinecite{Roy2017}. Throughout this work, we restrict ourselves to zero temperature.

The paper is outlined as follows. First, we describe the basic building block of our model, Majorana zero modes on mesoscopic superconducting islands, in Sec. \ref{majmes}.  Then, we describe the microscopic Hamiltonian of the system in Sec. \ref{model}.  We map the problem to coupled spins and rotors, with nearest-neighbor interactions using Jordan-Wigner transformation in Sec. \ref{sec_jw}. We compute the one-loop correction to the field theory in Sec. \ref{ft_loop_corr}. We present the conductivity calculations in Sec. \ref{cond}. We provide a concluding summary in Sec. \ref{concl}. Finally, Appendix \ref{mft} contains the mean-field calculation. 

\section{Superconducting islands in the topological regime}
\label{majmes}

In this section, we provide a concise summary of how charging effects can be
consistently incorporated in the mathematical description of topological
superconductors.  We consider a toy model for a topological superconductor,
the Kitaev wire.  \cite{Kitaev2000} It describes the effect of coupling a
quantum wire (modeled by a chain of spinless, fermionic modes) to an
infinitely large superconductor. As the superconductor screens the charge of
the electrons in the quantum wire, the relevant fermionic degrees at low
energies are chargeless (Majorana) fermions.\cite{Chamon2010, Beenakker2014}
Even more interesting, in the topological phase of the Kitaev wire, Majorana
zero modes, denoted by operators $\gamma_a, \gamma_b$, appear at its ends.
These operators are Hermitian $\gamma_j = \gamma_j^\dag$ and obey the Clifford
algebra $\{\gamma_i, \gamma_j\} = 2 \delta_{ij}$. Furthermore, the Majorana
zero modes lead to a twofold degeneracy of the ground state.  In a topological
superconductor, these exotic degrees of freedom are in fact non-Abelian
particles.\cite{Read2000,Ivanov2001} As a result, topological superconductors
realize a platform for topological quantum
computation.\cite{Nayak2008,Beenakker2013}

All these results essentially rely on the fact that the superconducting phase
$\phi$ is a classical variable that has a well-defined value.  The problem of
extending the results to a superconducting island with a fluctuating phase
$\phi$ has been addressed, \emph{e.g.}, in
Refs.~\onlinecite{Fu2010,Xu2010,VanHeck2011}. It has been shown that the
connection between the neutral Majorana zero mode operators $\gamma_a,
\gamma_b$ and the charged electronic annihilation operator $c$ is given by $c
= e^{-i\phi/2} (\gamma_a + i \gamma_b)/2$.\cite{Alicea2012} Note that the
charge on the superconducting island is carried by the Cooper-pairs in the
condensate and thus, is measured by $Q = -2 e i (d/d\phi)$ where $e>0$ is the
elementary charge.  Thus, the process of removing an electron from the
superconducting island (described by $c$)  both flips the fermion parity
$\mathcal{P} = -i \gamma_a \gamma_b$ (because of the action of $\gamma_a +i
\gamma_b$) and removes a charge $-e$ from the island (because of the action of
$e^{-i\phi/2}$). After all, fundamentally, the particles on the topological
superconductor are electrons.  So any physical state of the system has to be
reachable by applying the electronic operators $c$, $c^\dag$ to the vacuum
state without any particles present.

These arguments lead to the conclusion that the physical Hilbert space is not
simply given by the tensor product $\mathcal{H}^\otimes=\mathcal{H}_\phi \otimes
\mathcal{H}_{\gamma}$ of the condensate Hilbert space $\mathcal{H}_\phi$ (on
which $\phi$ and $Q$ act) and the Majorana Hilbert space $\mathcal{H}_\gamma$
(on which $\gamma_a$ and $\gamma_b$ act). Instead, the states $\psi \in
\mathcal{H}^\otimes$ are physical only if they are connected to the
vacuum state by addition or removal of electrons. Counting the fermion parity, denoted by the operator $\mathcal{P}$, with the charge on the island, denoted by the operator $Q$, yields the constraint
\begin{equation}\label{eq:fermion_constraint}
	(-1)^{Q/e} \psi = \mathcal{P} \psi.
\end{equation}
All states in the physical Hilbert space have to fulfill this constraint. Therefore, in the case of
a pair of Majorana zero modes on an island, even though the zero modes are chargeless, the two ground
states $|\pm\rangle$ (that differ in fermion parity with
$\mathcal{P}|\pm\rangle = \pm |\pm\rangle$) are not degenerate in energy. This is because 
they necessarily correspond to a different charging of the mesoscopic superconducting island:
$Q \in\{ 0, 2 e, 4e ,\cdots\}$ for the case of $\mathcal{P}=+$ and  $Q \in\{
e, 3 e, 5e ,\cdots\}$ for the case of $\mathcal{P}=-$.

On first sight, it might appear that the finite charging energy of a mesoscopic 
superconducting island  prevents any degeneracy due to the neutral Majorana
zero modes. However, it can be easily seen that this conclusion is incorrect
for the situation where there are two Kitaev wires and thus four Majorana zero
modes $\gamma_a, \gamma_b, \gamma_c, \gamma_d$ on the island. In this
case, the fermion parity constraint still is given by
\eqref{eq:fermion_constraint} but with a  fermion parity operator
$\mathcal{P} =  -\gamma_a \gamma_b \gamma_c
\gamma_d$ that involves all of the zero modes.  In
particular, we imagine to pair  $\gamma_a, \gamma_b$ ($\gamma_c, \gamma_d$)
with the ground state degeneracy given by the two eigenvalues of the partial
parity $\mathcal{P}_{ab} = i \gamma_a \gamma_b$ ($\mathcal{P}_{cd} = i
\gamma_c \gamma_d$). In the Majorana sector, the ground state is four fold
degenerate with the four states given by  $ |p_1, p_2\rangle =
|p_1\rangle_{ab} \otimes |p_2\rangle_{cd}$, with $p_1, p_2 \in \{\pm\}$. As
before, the finite charging energy will split states that correspond to
different eigenvalues of the charge operator $Q$. However, in this case the
constraint \eqref{eq:fermion_constraint} only implies that the states with
different total fermion parity $\mathcal{P}$ are not degenerate. This leaves
the degeneracy of the set of states $\{|+,+\rangle , |-,-\rangle\}$  and of
the set of states $\{|+,-\rangle , |-,+\rangle\}$. So even at finite charging
energy, there is a finite topological degeneracy of the ground state
remaining.\footnote{Note that for case of $2n$ Majorana zero modes, the
remaining topological degeneracy will be $2^{n-1}$.}

Physically, the degeneracy of the states $\{|+,-\rangle , |-,+\rangle\}$ can
be understood due to the fact that a single unpaired electron can
be moved from the fermionic mode $ab$ to $cd$ (or the other way round). As a result, the
fermion parity on each of the two pairs changes without changing the charging
energy. More interestingly, the degeneracy of the state $|+,+\rangle$ with
$|-,-\rangle$ arises from the fact that starting with $|+,+\rangle$ the state
$|-,-\rangle$ can be reached by breaking a Cooper pair and filling both the modes $ab$ and $cd$. This process does not involve taking any charge to or from the island and thus, does not change the charge $Q$ on the superconducting island. Note that due to the presence of the zero modes, one does not pay the usual superconducting energy gap to break the Cooper-pair. 

A remaining question is how to describe the tunneling of single charges
between two superconducting islands $i,j$. Note that in a multi-island
situation the parity constraint \eqref{eq:fermion_constraint} has to be
fulfilled on each of the islands separately. For concreteness, let us assume
that the zero modes $d$ on island $i$ is close to the mode $a$ on island $j$.
The resulting tunneling of single electrons leads to a term
\begin{equation}\label{eq:tunneling}
	H_{e\leftrightarrow e} \propto
	i\gamma^i_d \gamma^j_a \cos\left(\frac{\phi_i -\phi_j}{2}\right),
\end{equation}
also dubbed $4\pi$-periodic Josephson effect.\cite{Kitaev2000} It is easy to
see that under the action of this Hamiltonian [Eq. \eqref{eq:tunneling}], the state still satisfies the constraint [Eq. 
\eqref{eq:fermion_constraint}] both on island $i$ and $j$. The reason is that
while the $\cos$-term adjusts the charge (by increasing it by $e$ on one of the
islands and decreasing it on the other island correspondingly), the factor
$i\gamma^i_d \gamma^j_a$ flips the fermion parity on each side such that
physical states are mapped onto physical states in this process.

\section{Model Hamiltonian of the Majorana toric code}
\label{model}
\begin{figure} \includegraphics[width = .95\linewidth]{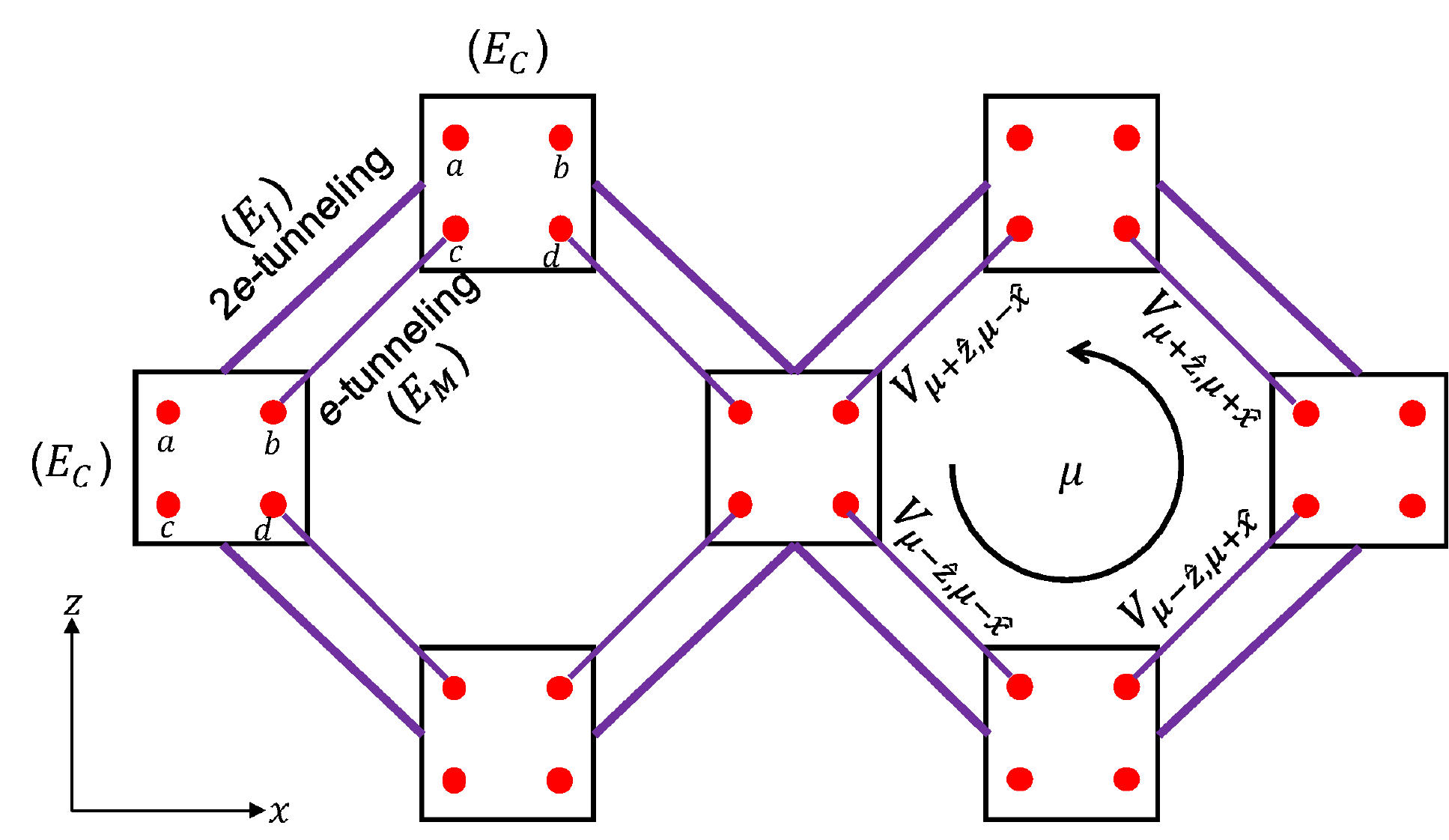}
\caption{\label{fig_1} (Color online) Schematic of two plaquettes of the lattice of
Majorana zero modes (denoted by red dots) on superconducting islands (denoted by
white squares).\cite{Roy2017} The three relevant energy scales are shown: the charging energy of each island $E_C$, the rate of tunneling of Cooper-pairs $E_J$ (denoted by thick links) and the rate of tunneling of single-electrons $E_M$ (denoted by thin links).}
 \end{figure}

With this introduction, it can be seen that the Hamiltonian of the system
depicted in Fig.~\ref{fig_1} is given by $H = H_C + H_J + H_M$, where 
\begin{align}\label{h0} 
  H_C &= 4E_C\sum_i n_i^2,\qquad H_J = -E_J\sum_{\langle
i,j\rangle} \cos(\phi_i-\phi_j), \nonumber\\ 
H_M &= -E_M \sum_{\langle
i,j\rangle}V_{i,j}\cos\Big(\frac{\phi_i-\phi_j}{2}\Big).  
\end{align} 
Here, the superconducting phase $\phi_i$ and the excess charge $n_i= Q_i/2e$ (in units of
Cooper pairs) on the $i$-th island are canonically conjugate. We
treat the idealized case of zero offset charges in the  absence of disorder.
The Majorana tunneling operator $V_{ij}$ between the two neighboring islands
$i,j$ is given by (see Fig.~\ref{fig_1}) \cite{Terhal2012} 
\begin{align}
\label{vij} V_{\mu+\hat{z}, \mu-\hat{x}} &= i\gamma_c^{\mu+\hat{z}}
\gamma_b^{\mu-\hat{x}},& V_{\mu-\hat{z},\mu-\hat{x}} &=
i\gamma_a^{\mu-\hat{z}}\gamma_d^{\mu-\hat{x}},\nonumber\\ V_{\mu-\hat{z},
\mu+\hat{x}} &= i \gamma_c^{\mu+\hat{x}}\gamma_b^{\mu-\hat{z}},&
V_{\mu+\hat{z},\mu+\hat{x}} &= i\gamma_a^{\mu+\hat{x}}\gamma_d^{\mu+\hat{z}},
\end{align} 
where the $\gamma^i_\alpha$ are Hermitian
operators. The fermion parity on the $i$-th island is given by the
operator  $\mathcal{P}_i = -\gamma^{i}_a\gamma^{i}_b\gamma^{i}_c\gamma^{i}_d$.
As the charge is constraint by the fermion parity, the (physical)
Hilbert-space for the Hamiltonian $H$ is spanned by the wavefunctions
satisfying $\psi(\phi_i+2\pi) = e^{i \pi Q_i/e} \psi(\phi_i)
=\mathcal{P}_i\psi(\phi_i)$ [cf. Eq. \eqref{eq:fermion_constraint}].\cite{Fu2010} At finite charging energy, the
ground state is in the even parity sector on each island (${\cal P}_i\equiv
+1$). In this sector, the four Majorana zero modes on each island encode one
qubit \cite{Bravyi2006} and, neglecting $H_J$, a perturbation calculation in
$E_M/E_C$ yields the toric code Hamiltonian.\cite{Xu2010, Landau2016}

\section{Mapping to a coupled spin-rotor Hamiltonian}
\label{sec_jw}
In this section, we map the Hamiltonian to that of spins coupled to rotors using Jordon-Wigner transformation.  Note that the Jordan-Wigner mapping presented in this section is different from the one presented in Ref.~\onlinecite{Terhal2012}. Unlike the mapping of Ref.~\onlinecite{Terhal2012}, the current one keeps the size of the Hilbert space invariant and thus, captures the degeneracies in the spectrum of the Hamiltonian. 

First, we perform  a gauge transformation in order to simplify the Hilbert-space to $2\pi$-periodic functions.\cite{VanHeck2012} This is done by applying a unitary transformation 
\begin{equation}
H\rightarrow \Omega^\dagger H \Omega,\ \psi\rightarrow \Omega^\dagger\psi
\end{equation}
where 
\begin{equation}
\Omega = \prod_i e^{iq_i\phi_i/2},\ q_i = \frac{1-{\cal P}_i}{2}.
\end{equation}
As a result, now only $2\pi$-periodic wavefunctions correspond to the physical states of the system. After the transformation, $H_J$ stays invariant, while the $H_C, H_M$ gets transformed as
\begin{align}
H_C &= 4E_C\sum_{i}\Big(n_i + \frac{q_i}{2}\Big)^2,\nonumber\\
H_M &= -\frac{E_M}{2}\sum_{\langle i,j\rangle}\Big(e^{-iq_i\phi_i}V_{i,j}e^{iq_j\phi_j}+{\rm{H.c.}}\Big).
\end{align} 

\begin{figure}
  \includegraphics[width = .95\linewidth]{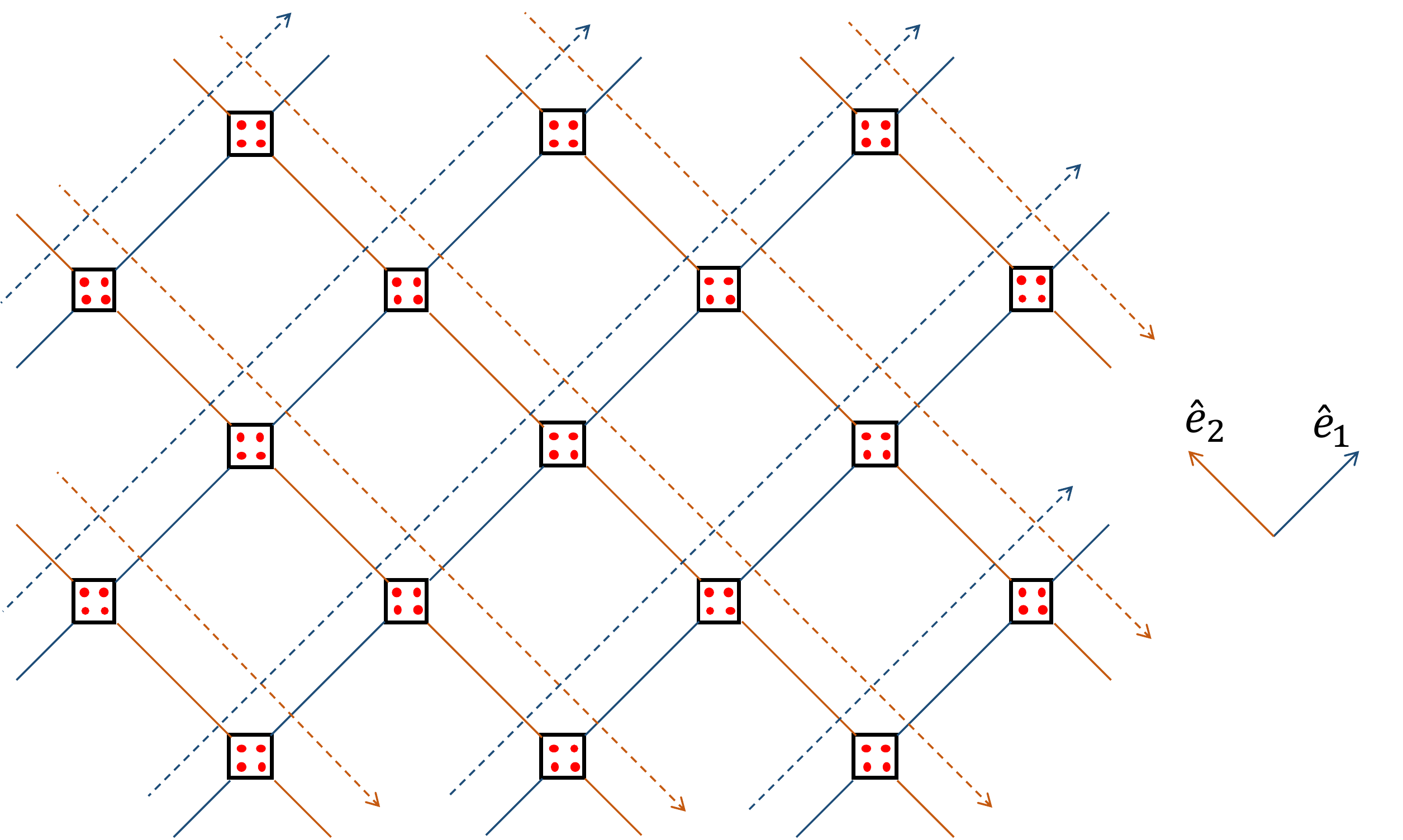}
\caption{\label{jwfig} (Color online) Schematic of the Jordan-Wigner transformation for the Majorana zero modes for an $L\times L$ lattice for both open and periodic boundary conditions.   }
\end{figure}

Next, we map the Majorana zero modes into spins using a Jordan-Wigner
transformation. Consider an $L\times L$ lattice with periodic boundary conditions. Starting
from any point on the lattice, we enumerate the Majorana zero modes traversing
the lattice first in the $\hat{e}_1$ direction (along the blue arrows in Fig.
\ref{jwfig}). Once all the $L$ lines along the $\hat{e}_1$ direction have been
traversed, we do the same in the $\hat{e}_2$ direction (along the orange
arrows in Fig.~\ref{jwfig}). Thus, each island is traversed twice, once in the
$\hat{e}_1$ and then in the $\hat{e}_2$ direction. For the $i^\text{th}$-island, while traversing in the $\hat{e}_{1(2)},$ direction, the map ${\cal M}$ maps the Majorana zero modes to spins as follows: 
\begin{align}
\label{jwmap}
{\cal M}(\gamma_{2i_k-1}) &= \Bigg[\prod_{j=1}^{i_k-1}\sigma^z_{j}\Bigg]\sigma^x_{i_{k}},\nonumber\\ {\cal M}(\gamma_{2i_{k}}) &= \Bigg[\prod_{j=1}^{i_k-1}\sigma^z_j\Bigg]\sigma^y_{i_k},\ k = 1,2.
\end{align}
Under this mapping, the fermion parity operator of the $i^\text{th}$-island is given by 
\begin{equation}
\mathcal{P}_i = -\gamma_a^i\gamma_b^i\gamma_c^i\gamma_d^i=\gamma_{2i_1-1}\gamma_{2i_1}\gamma_{2i_2-1}\gamma_{2i_2}=\sigma^z_{i_1}\sigma^z_{i_2}.
\end{equation}
Using Eqs.~\eqref{vij} and \eqref{jwmap}, for any link at the interior of the lattice (connecting sites $i,j$ with neither $i,j$ being a multiple of $L$), the $V_{i,j}$ are transformed as
\begin{equation}
V_{i_k,j_k} = \sigma^x_{i_k}\sigma^x_{j_k}, \ k=1,2.
\end{equation}
The interactions on the links at the boundary, where the lattice wraps around (connecting sites $i,j$ with $i$ or $j$ being a multiple of $L$) are nonlocal. For instance, along the first line of enumeration, the interaction coupling the Majorana modes $\gamma_{2L}$ and $\gamma_1$ is
\begin{equation}
V_{L,1} = \Bigg[\prod_{i=1}^L\sigma^z_i\Bigg]\sigma^x_1\sigma^x_L. 
\end{equation}
However, one can check that the product of the fermion parity along each of these lines is a conserved quantity.\cite{Nussinov2012} Thus, the Hilbert space splits into sectors where each of these  $\prod_{i=1}^L\sigma^z_i=\pm1$. In what follows, we will fix all the these quantities to be $+1$. Similar analysis can be done for other choices. Naturally, for open boundary conditions, this nonlocal interaction does not arise. Thus, the thermodynamic properties are described by the interactions on-site and those mediated by the internal links. 

\begin{figure}
  \includegraphics[width = .95\linewidth]{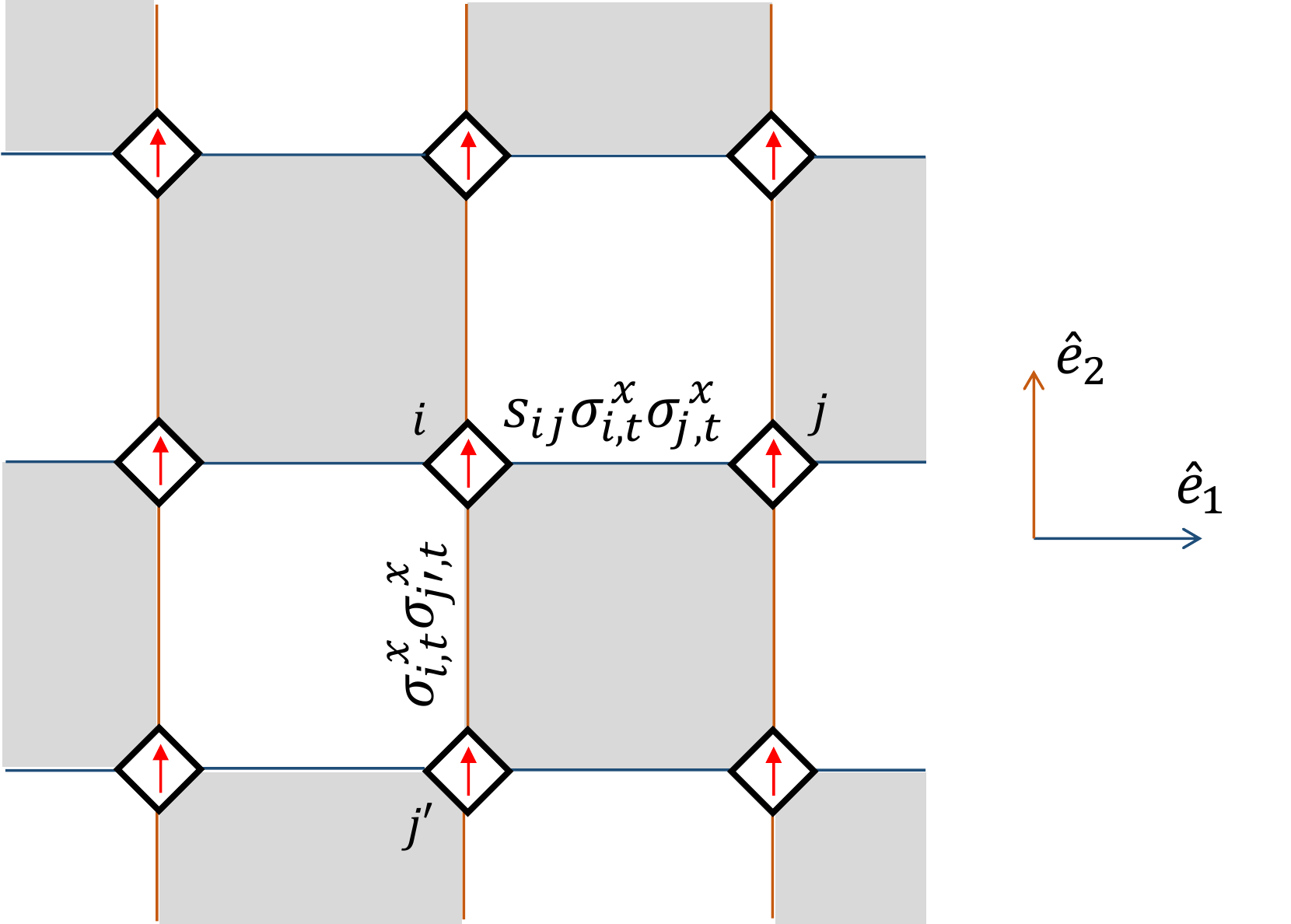}
\caption{\label{jwfig1} (Color online) Schematic of an inner $3\times3$ block
of the rotated lattice. The interaction between the
$i^\text{th}$, $j^\text{th}$ and the $i^\text{th}$, $j'{}^\text{th}$ islands are
shown after the Jordan-Wigner transformation in the Bell-basis. The
interaction on the horizontal link (in the $\hat{e}_1$ direction) connecting
the $i^\text{th}$ and the $j^\text{th}$ islands is given by
$s_{ij}\sigma^x_{i,t}\sigma^x_{j,t}$, where $s_{ij}$ is the product of the
$\sigma^z_s$ eigenvalues of the sign qubits on the corresponding islands. The
interaction on the vertical link (in the $\hat{e}_2$ direction) between the $i^\text{th}$ and the
$j^\text{th}$ islands is given by $\sigma^x_{i,t}\sigma^x_{j,t}$. }
\end{figure}

Next, we reduce the effective Hilbert space size that the interaction
Hamiltonian acts on. To that end, we lay down a Bell-basis for the spins on
the $i^\text{th}$-island.\cite{Terhal2012} We define a sign qubit ($s$) and a target qubit ($t$) on each island, whose joined state is given by $|\psi^{s,t}\rangle \equiv |s,t\rangle$: 
\begin{align}
|s=0,t=0\rangle &= \frac{1}{\sqrt{2}}(|00\rangle + |11\rangle),\nonumber\\ |s=0,t=1\rangle &= \frac{1}{\sqrt{2}}(|01\rangle + |10\rangle), \nonumber\\
|s=1,t=0\rangle &= \frac{1}{\sqrt{2}}(|00\rangle - |11\rangle),\nonumber\\ |s=1,t=1\rangle &= \frac{1}{\sqrt{2}}(|01\rangle - |10\rangle).
\end{align}
Thus, $s$ is the sign bit and the $t$ is the two-qubit parity bit of information in the superposition. In this basis, the operators $\sigma^x_{i_k}, \sigma^z_{i_k}$, $k=1,2$ get mapped to:
\begin{align}
\sigma^x_{i_1}|s,t\rangle = \sigma^x_{i,t}\sigma^z_{i,s}|s,t\rangle, \ \sigma^x_{i_2}|s,t\rangle = \sigma^x_{i,t}|s,t\rangle,\nonumber\\
\sigma^z_{i_1}|s,t\rangle = \sigma^x_{i,s}|s,t\rangle, \ \sigma^z_{i_2}|s,t\rangle = \sigma^z_{i,t}\sigma^x_{i,s}|s,t\rangle,
\end{align}
where $\sigma^x_{i,s(t)}, \sigma^z_{i, s(t)}$ are the Pauli operators for the
sign (target) qubits. Thus, the interaction Hamiltonian
$\sigma^x_{i_1}\sigma^x_{j_1}$ along $\hat{e}_1$ direction gets mapped to
an effective interaction between the target qubits where the sign of the
interaction is determined by the $\sigma^z_s$ eigenvalue of the sign qubits.
We can define a classical bit $s_{ij}$ on each link connecting the
$i^\text{th}$ and the $j^\text{th}$ island in the $\hat{e}_1$ direction that
is the product of the $\sigma^z$-eigenvalues of the corresponding sign qubits
(cf.\ Fig.~\ref{jwfig1}). 
The resultant interaction Hamiltonian, after a trivial rotation on the target qubits, is then written by (we drop the $t$-index on the spins for clarity)
\begin{align}
\label{hc}
H_C &= 4E_C\sum_i \Big(n_i+\frac{1+\sigma_i^z}{4}\Big)^2,\\
H_M &= -\frac{E_M}{2} \sum_{\langle
i,j\rangle}s_{i,j}\Big\{\sigma_i^-\sigma_j^-(e^{i\phi_i} +
e^{i\phi_j})\nonumber\\& \qquad\qquad\qquad
+\sigma_i^-\sigma_j^+\big[1+e^{i(\phi_i-\phi_j)}\big]
+ \text{H.c.}\Big\}, \nonumber
\end{align}
while $H_J$ remains invariant.\cite{Cooper_pair_note}  Here, the sign of the interaction is
determined by gauge bits $s_{i,j}=\pm1$. The state-space is subjected to the following constraints: 
\begin{equation}
\prod_i\sigma^z_i = 1, \ \prod_w\prod_\square s_{ij} = 1,
\end{equation}
where $\prod_w$ denote the product over all the white plaquettes (see Fig. \ref{jwfig1}).\footnote{Note that $\prod\prod_\square s_{ij} = 1$ is trivially true.} The above constraints arise from the fact that the total fermion number parity of all the islands is conserved. The spectrum of the Hamiltonian in Eq.~\eqref{h0} is given by the union of the spectra of Eq.~\eqref{hc}  for different gauge bit configurations. Here, gauge bit configurations are nonequivalent if their product around any plaquette differs.\cite{Fradkin2013} The Hamiltonian is invariant under the simultaneous transformations $U_\theta$:
\begin{equation}
\label{globsymm}
e^{i\phi_i}\mapsto e^{i\phi_i}e^{i\theta},\ \sigma_i^+\mapsto \sigma_i^+e^{i\theta/2}.
\end{equation}
Physically, this global symmetry originates from the fact that the spins
correspond to  single-electrons that carry half of the charge of the Cooper
pairs.  

The different phases of the system arise depending on how $H_C, H_M$ or $H_J$ break the $U(1)$ symmetry.\cite{Roy2017} In the phase when the Josephson tunneling is the strongest, the $U(1)$ symmetry is spontaneously broken, the rotors align and the ground state  of the system is invariant under rotation by multiples of $2\pi$. Then, the system is a conventional superconductor of Cooper pairs (of charge $2e$). We denote this phase by $\{2\pi\}$. When the Majorana-assisted single electron tunneling is the strongest, the $U(1)$ symmetry is again broken. However, now the spins are aligned and the ground state is invariant only under rotation by multiples of $4\pi$. Moreover, the alignment of the spins in turn aligns the rotors. As a result, the system is an `exotic' superconductor \cite{Wen1991,Balents1999, Senthil2000} of charge-$e$ bosons. We denote this phase by $\{4\pi\}$. Finally, when the charging energy is the strongest, the complete $U(1)$ is restored, the rotors and the spins are disordered and the system is a Mott insulator. We denote this phase by $\{\theta\}$. 

The relevant field theory capturing the phase-diagram of this system was proposed in Ref.~\onlinecite{Roy2017}. In the next section, we compute the one-loop correction to the field theory describing the transition between the phases $\{\theta\}$ and $\{2\pi\}$. Our calculation shows that an additional tricritical point is likely to be present in the phase-diagram which was not captured by the mean-field analysis of Ref.~\onlinecite{Roy2017}. 
\section{One loop correction to the field theory}
\label{ft_loop_corr}
Consider the coarse-grained expectation values of $e^{i\phi_i}$ and $\sigma^+_i$ in
imaginary time $\tau$, denoted by complex fields 
$\psi_r(\bm{r},\tau)$ and $\psi_s(\bm{r},\tau)$. The relevant degrees of freedom are given
by the low-frequency, long-wavelength behavior of these complex fields. 
The microscopic symmetry $U_\theta$ is
elevated to the symmetry $\psi_r\mapsto \psi_r e^{i\theta},
\psi_s\mapsto \psi_s e^{i\theta/2} $ on the coarse-grained
variable that has to be respected in the effective field theory. Since close to
the phase transitions, the fields are small, the action can be expanded in a Taylor and gradient
expansion in $\psi_r, \psi_s$.
 The partition function at zero temperature is given by $Z = \int\! {\cal
D}\psi_s\,{\cal D}\psi_s^*\,{\cal D}\psi_r\,{\cal
D}\psi_r^*\,e^{-S}$ with the Euclidean action
\begin{align}\label{action_tot}
  S= & \int d^2r\, d\tau \Bigl[  
    |\partial_\tau\psi_s|^2 + |\partial_\tau\psi_r|^2  
    + K_s|\nabla\psi_s|^2+K_r|\nabla\psi_r|^2
    \nonumber \\
   &+r_M |\psi_s|^2 + 
   r_J |\psi_r|^2  + u_s|\psi_s|^4+
   u_r|\psi_r|^4 
   \nonumber\\
   &+\beta|\psi_s|^2|\psi_r|^2 - \alpha({\psi_s^*}^2\psi_r +
 \psi_s^2\psi_r^*)  \Bigr].
\end{align}
The parameter $r_x$ is used to tune through the phase transition and corresponds to $-E_x/E_C$, where $x = M,J$. From stability considerations, $u_s, u_r, \beta$ must be positive and we choose $\alpha>0$ without loss of generality. 

The phase transition between the phases $\{\theta\}$ and
$\{2\pi\}$ is the conventional Mott-insulator to superconductor transition.\cite{Fisher1989, Herbut2007,
Sachdev2011}  Across this transition, $\psi_s$ stays zero,
while $|\psi_r|$ turns finite. Integrating over small
fluctuations $\delta\psi_s$ around the saddle point $\bar{\psi}_s = 0$, we get an
effective partition function $Z^{(b)} = \int\! {\cal D}\psi_r\,{\cal
D}\psi_r^*\, e^{-S^{(b)}}$ with\cite{Roy2017}
\begin{align}
\label{sb}
S^{(b)} &= \int\! d^2r\, d\tau\big(|\partial_\tau\psi_r|^2+K_r|\nabla\psi_r|^2
+ r_J |\psi_r|^2 + u_r|\psi_r|^4\big)\nonumber.
\end{align}
Here, we have kept the renormalization of the couplings to zeroth order in $\alpha, \beta$. The phase transition line is given by $r_J=0$ and it becomes metastable in the phase $\{4\pi\}$. Next, we perform calculations going beyond the zeroth order in $\alpha, \beta$. We show that due to the cubic term proportional to $\alpha$ in Eq. \eqref{action_tot}, this line of phase transition also terminates in a tricritical point. After this point, the transition turns first order. For the ease of computation, we define $\mathbf{x}=(\mathbf{r},\tau)$ and rescale the spatial axis so as to set $K_s = 1$. The rescaled stiffness of the rotor sector is denoted by $\tilde{K}_r$.

\begin{figure}[H]
  \includegraphics[width = .95\linewidth]{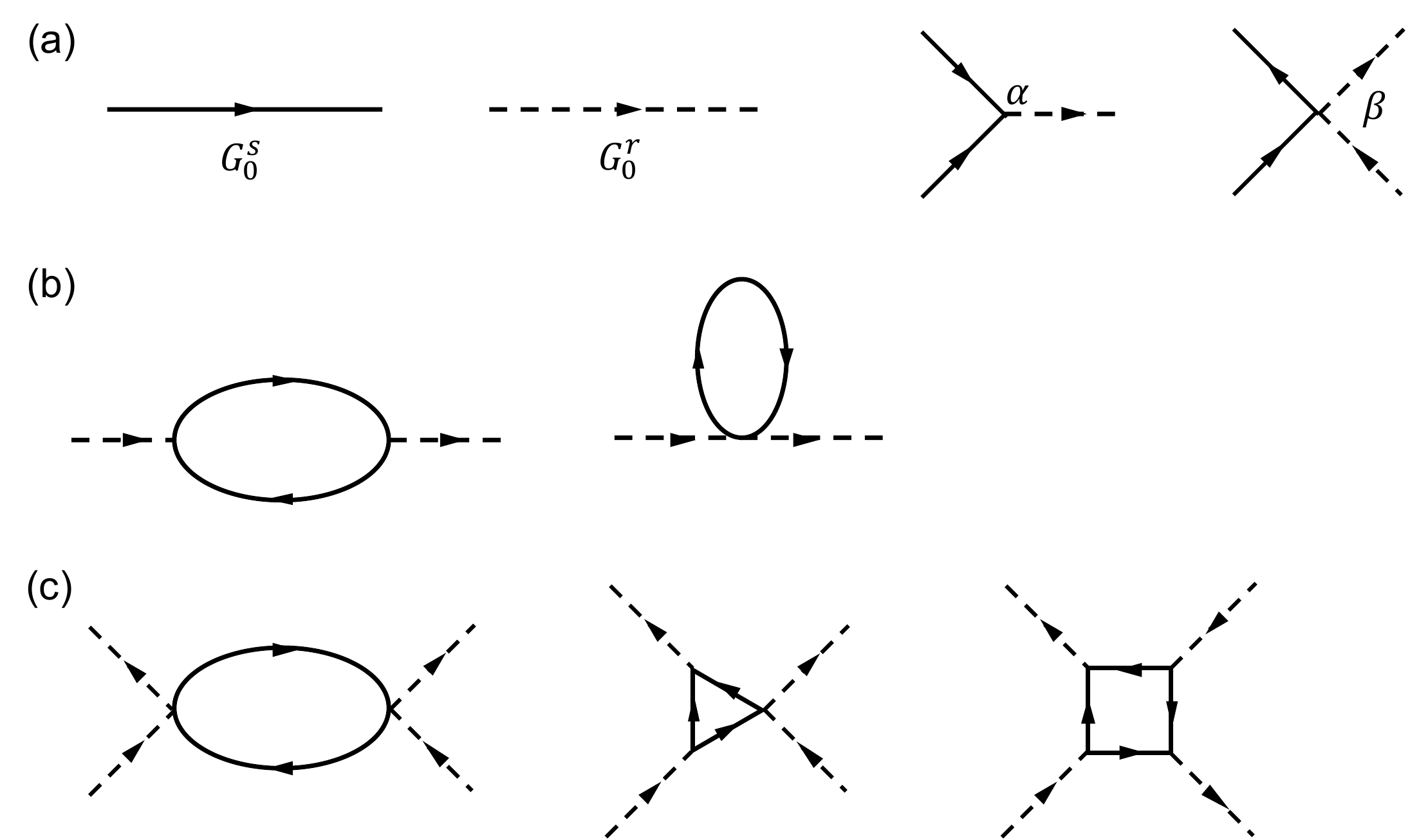}
\caption{\label{fig_loop1} (a) The solid (dashed) line corresponds to the propagator for the $\psi_{s(r)}$ field. The relevant nonlinear interaction vertices coupling $\psi_s, \psi_r$ are shown. The three (four) point vertex corresponding to the cubic (quartic) interaction with coupling strength $\alpha(\beta)$. (b) Diagrams contributing to the renormalization of the propagator of the $\psi_r$ field. (c) Diagrams contributing to the renormalization of the interaction term $u_r |\psi_r|^4$.  }
\end{figure}

\begin{figure}
  \includegraphics[width = .95\linewidth]{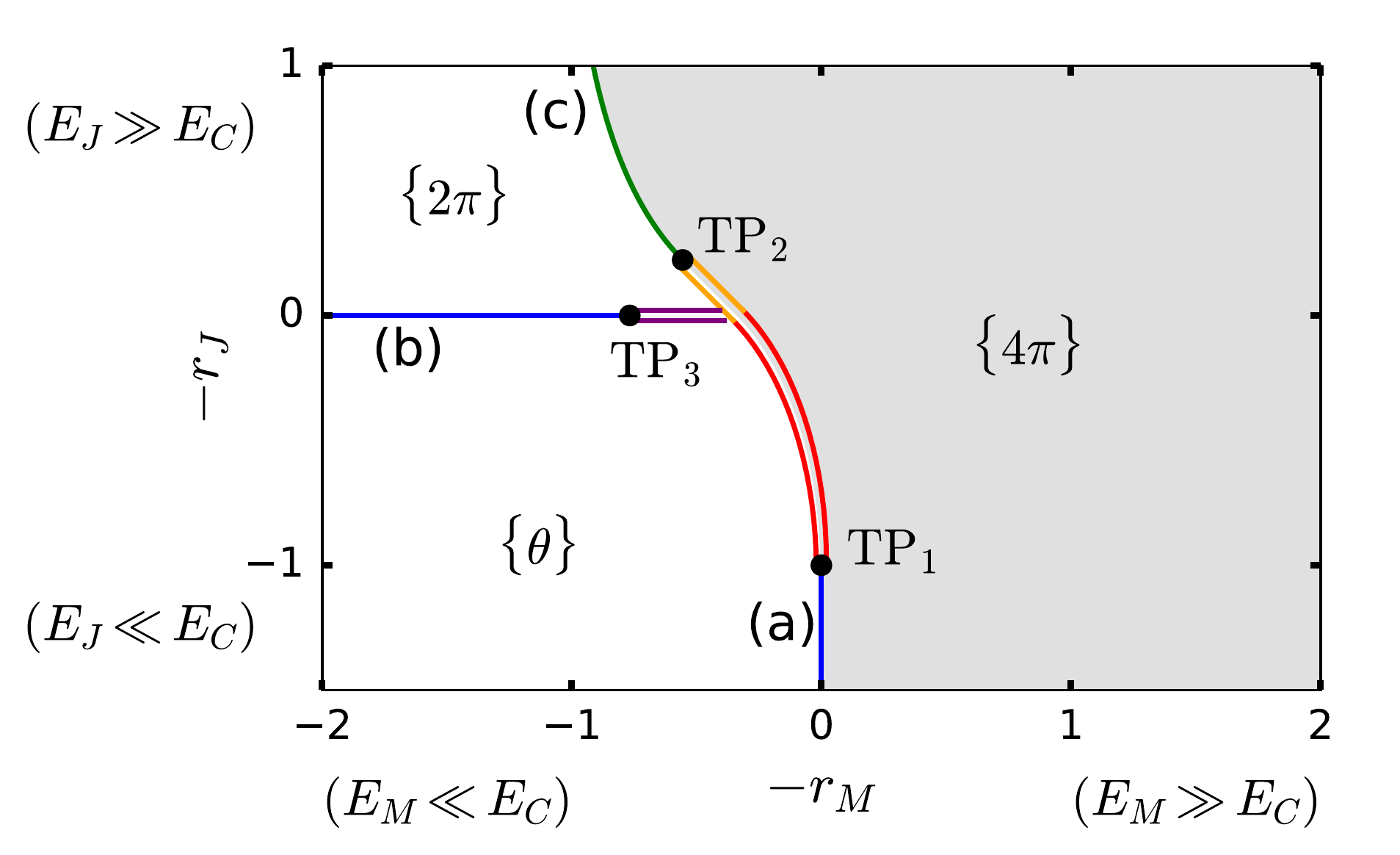}
\caption{\label{pd} (Color online) Phase diagram for the model\cite{Roy2017} as a
function of $r_J,r_M$ for $u_\text{s}=u_\text{r}=\alpha=\beta=1$. The blue
line, marked with (a), denotes a (2+1)D-XY phase transition line, separating
the phases $\{\theta\}$ and $\{4\pi\}$. This transition terminates
in a tricritical point ($\rm{TP_1}$), after which the transition becomes a
first-order transition (shown as red double line).  The blue line, marked with
(b), also denotes a (2+1)D-XY phase transition line, that separates the phases
$\{\theta\}$ and $\{2\pi\}$. This transition line terminates at the tricritical point $\rm{TP_3}$, after which the transition turns first order (shown as purple double line). The tricritical point $\rm{TP_3}$ is predicted from the one-loop calculations of this work and was missed in the earlier mean-field predictions of Ref. \onlinecite{Roy2017}. The first order line out of $\rm{TP_3}$ meets the first-order line coming out of $\rm{TP_1}$. The green line, marked with (c),
denotes a (2+1)D-Ising transition line separating the phases $\{2\pi\}$ and
$\{4\pi\}$. This phase transition line terminates in a tricritical point
($\rm{TP_2}$), after which turns into a first-order line (shown as orange
double line), which smoothly transforms into the red first-order line.}
\end{figure}

Considering small-fluctuations around the saddle point: $\psi_s = \bar\psi_s + \delta\psi_s$ and expanding to second order, the effective action can be written as
\begin{equation}
\label{sb1}
S^{(b)} =S[\delta\psi_s] + S[\psi_r] + S[\delta\psi_s, \psi_r]
\end{equation}
where the actions for the spin and rotor fields are given by:
\begin{align}
S[\delta\psi_s] &= \int_x\Big\{|\partial_\tau\delta\psi_s|^2 + |\nabla\delta\psi_s|^2 + r_M|\delta\psi_s|^2\Big\},\nonumber\\
S[\psi_r] &= \int_x\Big\{|\partial_\tau\psi_r|^2 + \tilde{K}_r|\nabla\psi_r|^2 + r_J|\psi_r|^2 +u_r|\psi_r|^4\Big\},\nonumber
\end{align}
while the interaction between the two is given by
\begin{equation}
S[\delta\psi_s, \psi_r] = \int_x\Big\{-\alpha(\delta\psi_s^{*2}\psi_r + \delta\psi_s^2\psi_r^*) +\beta|\delta\psi_s|^2|\psi_r|^2\Big\}.\nonumber
\end{equation}
Here, $\int_x\equiv\int d^3x$. Now, the field $\delta\psi_s$ can be
perturbatively integrated out. The diagrams contributing to the
renormalization of the action to one loop order are shown in Fig.~\ref{fig_loop1}. 
Thus, the effective action for the rotor field is given by
\begin{align}
S[\psi_r] &= \int_x\Big\{(1+\delta\tilde{K}_r)|\partial_\tau\psi_r|^2 + (\tilde{K}_r+\delta\tilde{K}_r)|\nabla\psi_r|^2\nonumber\\&\qquad + (r_J+\delta r_J)|\psi_r|^2 +(u_r+\delta u_r)|\psi_r|^4\Big\},\nonumber
\end{align}
where 
\begin{align}
\delta \tilde{K}_r &= 4\alpha^2\int_p\frac{1}{(r_M + p^2)^3}\nonumber\\
\delta r_J &= -4\alpha^2\int_p\frac{1}{(r_M + p^2)^2}+\beta \int_p\frac{1}{(r_M + p^2)}\nonumber\\
\delta u_r &= -\frac{\beta^2}{2}\int_p\frac{1}{(r_M + p^2)^2} + 4\alpha^2\beta \int_p\frac{1}{(r_M + p^2)^3}\nonumber\\&\quad-144\alpha^4\int_p\frac{1}{(r_M + p^2)^4},
\end{align}
where $\int_p = \int\frac{d^3p}{(2\pi)^3}$. Note that there are no infrared singularities in these diagrams since $r_M\neq0$ as we are far from the phase-transition in the spin-sector. The renormalization of the spin-wave stiffness $\delta\tilde{K}_r$ can be evaluated to be: 
\begin{equation}
\delta\tilde{K}_r = \frac{1}{8\pi}\frac{\alpha^2}{r_M^{3/2}}
\end{equation}
The renormalization of the gap-parameter $\delta r_J$ has the usual cut-off-dependent shift as is common in one-loop renormalization group analysis:
\begin{equation}
\delta r_J = -\frac{\alpha^2}{2\pi r_M^{1/2}}-\frac{\beta r_M^{1/2}}{4\pi}+\frac{\beta r_M^{1/2}}{2\pi^2}\Lambda,
\end{equation}
where $\Lambda$ is the ultraviolet cut-off. Finally, the renormalization of $u_r$ is given by:
\begin{align}
\delta u_r &= -\frac{1}{16\pi}\frac{\beta^2}{r_M^{1/2}} + \frac{1}{8\pi}\frac{\alpha^2\beta}{r_M^{3/2}}-\frac{9}{4\pi}\frac{\alpha^4}{r_M^{5/2}}.
\end{align}
For large $\alpha$, $\delta u_r<0$. Also, for both large and small values of $r_M$, $\delta u_r<0$. 
Thus, without fine-tuning, for generic parameter choices, $\delta u_r<0$. This implies that the stabilizing quartic interaction is depressed. Thus, it is likely that this line of 3D-XY phase-transition also terminates in a tricritical point. Then, a sextic term generated from integrating out the $\psi_s$ field stabilizes the theory. After the tricritical point, the transition turns first order. The phase-diagram\cite{Roy2017}, including the additional tricritical point $\rm{TP_3}$, is shown in Fig. \ref{pd}. 

\section{Conductivities at the different phase-transitions}
\label{cond}
In this section, we investigate the charge-response signatures of the different continuous phase-transitions of our model. As discussed in the previous section, the transition between the phases $\{\theta\}$ and $\{2\pi\}$ is of 3D-XY type, which likely terminates in a 3D-XY tricritical point. The same was shown to be true for the transition between the phases $\{\theta\}$ and $\{4\pi\}$.\cite{Roy2017} The universal value of conductivity across the 3D-XY phase-transition lines has been calculated analytically using 1/N expansion \cite{Cha1991} and $\epsilon$ expansion\cite{Fazio1996}. These results can be directly applied for the 3D-XY transitions of our model and the 3D-XY tricritical points. On the other hand, a 3D-Ising line terminating at a 3D-Ising tricritical point separates the the phases $\{2\pi\}$ and $\{4\pi\}$. The charge response across this transition has not been analyzed before. We do this calculation using 1/N expansion.

In what follows, first, we summarize the results for the 3D-XY transitions. Since we need the 1/N formalism for the subsequent calculation, we sketch the details in the context of the 3D-XY transitions. Then, we perform the relevant analysis for the 3D-Ising transition. Finally, we provide the conductivities across the different tricritical points. 

The Kubo formula for the conductivity is given by\cite{Mahan2013, Sachdev2011}
\begin{equation}
\label{sigmaxx}
\sigma_{xx}(ik_z) = \frac{\hbar}{k_z}\int d^3x e^{i\mathbf{k}\cdot\mathbf{x}}\frac{\delta^2F}{\delta A_x(\mathbf{x})\delta A_x(0)}\Big|_{\mathbf{A}=0},
\end{equation}
where $\mathbf{x}=(\mathbf{r},\tau)$, $\mathbf{k} = (0,0,k_z)$ and $F=-\ln Z$. The coupling to the electromagnetic field is performed by making the previously encountered global U(1) symmetry, denoted by angle $\theta$ [cf. Eq. \eqref{globsymm}], a local one, $i.e.,$ space-time dependent: $\theta\rightarrow\theta(\mathbf{r},\tau)$. Thus, $Z$ can be obtained from the earlier action given by Eq.~\eqref{action_tot} after the minimal substitution:
$\nabla\psi_{s,r}\rightarrow
(\nabla-i\frac{e_{s,r}}{\hbar}\mathbf{A})\psi_{s,r}$. Here, $e_{s}$ and $e_{r}$ are the
charges of the spin and rotor fields and are given by $e$ and $2e$ respectively. This is because while the rotor sector conducts electric current via Cooper pairs, the spin sector, arising out of the Majorana zero modes, conduct current through single electrons. The DC-conductivity is obtained by first analytically continuing our result to $ik_z\rightarrow \omega+i0^+$, followed
by taking the limit of $\omega\rightarrow0$ (see Chap. 3 of Ref.~\onlinecite{Mahan2013}). The electric current is carried by both spin and rotor current fields, given by
\begin{align}
\label{currentdef}
\mathbf{J}_{s(r)}(\mathbf{x})&=\frac{1}{2i}\big\{\psi_{s(r)}^*(\mathbf{x})\nabla\psi_{s(r)}(\mathbf{x})-\psi_{s(r)}(\mathbf{x})\nabla\psi_{s(r)}^*(\mathbf{x})\big\}.
\end{align}
From Eq.~\eqref{sigmaxx}, we get: 
\begin{equation}
\label{sigmadef}
\sigma_{xx}(ik_z) = \frac{e_s^2}{\hbar}\frac{\rho_s(k_z)}{k_z} + \frac{e_r^2}{\hbar}\frac{\rho_r(k_z)}{k_z}-\frac{4e_se_r}{\hbar}\frac{\rho_{sr}(k_z)}{k_z},
\end{equation}
where
\begin{align}
\label{rhodef}
\rho_{s(r)}(k_z)&=2\langle\psi^*_{s(r)}(0)\psi_{s(r)}(0)\rangle\nonumber\\&\quad-4\int d^3x \langle \mathbf{J}_{s(r),x}(\mathbf{x}) \mathbf{J}_{s(r),x}(0)\rangle e^{i\mathbf{k}\cdot\mathbf{x}},\\\rho_{sr}(k_z)&=\int d^3x \big\{\langle \mathbf{J}_{s,x}(\mathbf{x}) \mathbf{J}_{r,x}(0)\rangle\nonumber\\&\quad+\langle \mathbf{J}_{r,x}(\mathbf{x}) \mathbf{J}_{s,x}(0)\rangle\big\} e^{i\mathbf{k}\cdot\mathbf{x}}.
\end{align}

\subsection{Conductivities at the 3D-XY transitions}
Consider the 3D-XY transition between the phases $\{\theta\}$ and $\{4\pi\}$ [line (a) in Fig. \ref{pd}]. The relevant action is given by\cite{Roy2017}
\begin{align}
\label{sa}
S^{(a)}&=\int d^3x\Big\{|\partial_\tau\psi_s|^2+K_s|\nabla\psi_s|^2 +
r_M |\psi_s|^2\nonumber\\&\quad
+ \frac{u_M}{2}|\psi_s|^4 +
\tilde{u}_M|\psi_s|^6 \Big\}.
\end{align}
For the $u_M>0$, the transition is of the 3D-XY type and the sextic term can be neglected. \cite{Chaikin2000} We will treat this case in this subsection. The tricritical points are treated in Sec. \ref{tps}. We perform perturbation in the inverse of the dimensionality of the order parameter $N$.\cite{Ma1973, Coleman1988, Cha1991, ZinnJustin2002} For this case, with complex
order parameter, $N=2$. The
effective action in terms of the N-component order parameter is given by
\begin{eqnarray}
\label{sa1}
S_{\rm{eff}}^{(a)}&=&\int d^3x\Big\{\nabla\psi_{s,\alpha}^*\nabla\psi_{s,\alpha} + r_M \psi_{s,\alpha}^*\psi_{s,\alpha}\nonumber\\&&+ \frac{u_M}{2}(\psi_{s,\alpha}^*\psi_{s,\alpha})^2\Big\},
\end{eqnarray}
where $\nabla$ now denotes the 3D gradient. We have rescaled the spatial axis so as to set $K_s = 1$. We consider $N'$ complex fields, $\psi_{s,\alpha}$, $\alpha=1,\ldots,N'$, where $N=2N'$.  

For this effective field theory, the conductivity is given by
\begin{equation}
\sigma_{xx}(ik_z) = \frac{e_s^2}{\hbar}\frac{\rho_s(k_z)}{k_z}, 
\end{equation}
where $\rho_s(k_z)$ is defined in Eq.~\eqref{rhodef} after the transformation:
$\psi_s\rightarrow \psi_{s,\alpha}$. The current operator is defined by Eq.~\eqref{currentdef} after the same transformation. We define the Fourier transform of $\psi_{s,\alpha}$:
\begin{equation}
\label{fourierdef}
\psi_{s,\alpha}(\mathbf{x}) \equiv \frac{1}{\sqrt{\Omega}}\sum_{\mathbf{q}}e^{i\mathbf{q}\cdot\mathbf{x}}\psi_{s,\alpha}(\mathbf{q}),
\end{equation}
where $\Omega$ is the 3D volume element.
In Fourier domain, the effective action is given by
\begin{align}
\label{sa2}
S_{\rm{eff}}^{(a)}&=\sum_{\mathbf{q}} (\mathbf{q}^2+r_M)\psi_{s,\alpha}^*(\mathbf{q})\psi_{s,\alpha}(\mathbf{q})+\frac{1}{2\Omega}\sum_{\mathbf{q}}u_M\nonumber\\&\qquad\sum_{\mathbf{p},\mathbf{p'}} \psi_{s,\alpha}^*(\mathbf{p}+\mathbf{q})\psi^*_{s,\beta}(\mathbf{p}')\psi_{s,\alpha}(\mathbf{p})\psi_{s,\beta}(\mathbf{p}'+\mathbf{q})\Big\}. 
\end{align}
The expression for $\rho_s(k_z)$ is given by:
\begin{widetext}
\begin{align}
\rho_s(k_z)= \frac{2}{\Omega}\sum_{\mathbf{q}} \langle \psi^*_{s,\alpha}(\mathbf{q})\psi_{s,\alpha}(\mathbf{q})\rangle-\frac{4}{\Omega}\sum_{\mathbf{p},\mathbf{q}}q_xp_x\langle \psi_{s,\alpha}^*(\mathbf{q}-\mathbf{k}/2)\psi_{s,\alpha}^*(\mathbf{p}+\mathbf{k}/2)\psi_{s,\alpha}(\mathbf{p}-\mathbf{k}/2)\psi_{s,\alpha}(\mathbf{q}+\mathbf{k}/2)\rangle.
\end{align}
\end{widetext}
Since in three dimensions, the quartic coupling is relevant, we treat it
perturbatively using the 1/N-expansion. To obtain a non-trivial result, we
must choose $u_M = g_M/N$, where $g_M$ stays constant. Following Ref.~\onlinecite{Coleman1988}, we can decouple the interaction by Hubbard-Stratonovich transformation, introducing an auxiliary field $\zeta(\mathbf{r})$, leading to an effective action:  
\begin{align}
\label{sa3}
S_{\rm{eff}}^{(a)}[\psi_{s,\alpha},\zeta]&=\int d^3x\Big\{\nabla\psi_{s,\alpha}^*\nabla\psi_{s,\alpha} + r_M \psi_{s,\alpha}^*\psi_{s,\alpha}\nonumber\\&\quad+ \frac{N}{2g_M}\zeta^2+i\zeta\psi_{s,\alpha}^*\psi_{s,\alpha}\Big\}.
\end{align}
The linear term in the above action can be removed by shifting the $\zeta$ field: $\zeta\rightarrow \zeta + \bar\zeta$, where $\zeta$ is determined by 
\begin{equation}
\frac{\partial \tilde{S}_{\rm{eff}}^{(a)}[\zeta]}{\partial\zeta}\Big|_{\zeta=\bar\zeta}=0, 
\end{equation}
with 
\begin{equation}
e^{-\tilde{S}_{\rm{eff}}^{(a)}}[\zeta] = \int {\cal D}\psi_{s,\alpha}{\cal D}\psi^*_{s,\alpha}e^{-S_{\rm{eff}}^{(a)}[\psi_{s,\alpha},\zeta]}.
\end{equation}
The action then becomes
\begin{align}
\label{sa4}
S_{\rm{eff}}^{(a)}[\psi_{s,\alpha},\zeta]&=\int d^3x\Big\{\nabla\psi_{s,\alpha}^*\nabla\psi_{s,\alpha} + \tilde{r}_M \psi_{s,\alpha}^*\psi_{s,\alpha}\nonumber\\&\quad+ \frac{N}{2g_M}\zeta^2+i\zeta\psi_{s,\alpha}^*\psi_{s,\alpha}+\frac{N}{g_M}\bar\zeta\zeta\Big\},
\end{align}
where $\tilde{r}_M = r_M + i\bar\zeta$, is the renormalized gap parameter that goes to zero at the phase-transition. The last two terms cancel, thereby eliminating the linear term in $\zeta$. The free propagator for the $\psi_{s,\alpha}$ field is given by
\begin{equation}
G^\psi_0(\mathbf{q}) = \frac{1}{\mathbf{q}^2+\tilde{r}_M} 
\end{equation}
and the resummed two-body interaction mediated by the $\zeta$ is given by
\begin{equation}
\tilde{u}_M(\mathbf{k}) = \frac{u_M}{1+u_MN'\Pi^\psi(\mathbf{k})},
\end{equation}
where 
\begin{equation}
\Pi^\psi(\mathbf{k}) = \int\frac{d^3q}{(2\pi)^3}G^\psi_0(\mathbf{k}+\mathbf{q})G^\psi_0(\mathbf{q}).
\end{equation}
Note that as $N'\rightarrow\infty$, the effective action is free with
$\tilde{u}_M\rightarrow0$. The 1/N-expansion can now be performed by replacing
$r_M, u_M$ by $\tilde{r}_M, \tilde{u}_M(\mathbf{k})$. The details of the
calculation can be found in Ref.~\onlinecite{Cha1991}. Here, we merely state the result: 
\begin{equation}
\sigma_{xx}=\frac{\pi}{8}\Big(1-\frac{1}{N'}\frac{32}{9\pi^2}\Big)\frac{e^2}{h}=\frac{\pi}{8}\Big(1-\frac{32}{9\pi^2}\Big)\frac{e^2}{h},
\end{equation}
where in the last equality, we have set $N'=1$ for the 3D-XY transition. 

For the phase-transition between the $\{\theta\}$ and the $\{2\pi\}$ phases [line (b) in Fig. \ref{pd}], the same calculation can be performed, with the following substitution: $e_s\rightarrow e_r=2e$. Thus, in that case, the DC-conductivity is given by
\begin{equation}
\label{sigmaval1}
\sigma_{xx}=\frac{\pi}{2}\Big(1-\frac{32}{9\pi^2}\Big)\frac{e^2}{h}.
\end{equation}

\subsection{Conductivity at the 3D-Ising transition}
For the effective action for the transition between $\{2\pi\}$ and $\{4\pi\}$ [line (c) in Fig. \ref{pd}], we use the
parametrization\cite{Roy2017}
\begin{eqnarray}
\label{def1}
\psi_r &=& (\bar{\rho}_r + \delta\rho_r)e^{i\theta_r/\bar{\rho}_r},\\\label{def2}
\psi_s &=& (\sigma+iw)e^{i\theta_r/2\bar{\rho}_r}
\end{eqnarray}
where
$\bar{\rho}_r$ is the saddle point value of $|\psi_r|$ and the
real fields $\delta \rho_r$, $\theta_r$ and $\sigma$, $w$ denote
the fluctuations of $\delta\psi_r$ and $\delta\psi_s$.  The
fluctuations in $\theta_r$ correspond to the massless Goldstone mode
associated with the symmetry breaking in the rotor sector. They decouple from
the rest.  As described in Ref. \onlinecite{Roy2017}, the emergent Ising degree of freedom $\sigma$ undergoes the transition, while the fields $w, \delta\rho_r$ stay regular with gap parameters $r_w, r_{\delta\rho_r}>0$. The effective action for the Ising degree of freedom is given by
\begin{eqnarray}
\label{sc}
S^{(c)} &=  \int d^3x\big\{(\partial_\tau\sigma)^2+K_s(\nabla\sigma)^2 +t_c \sigma^2
  + \frac{u_c}{2}\sigma^4\nonumber\\& \quad+\!\tilde{u}_c\sigma^6
 \big\}. 
\end{eqnarray}
Again, in this subsection, we consider the case when $u_c>0$ (see Sec. \ref{tps} for $u_c=0$). In this new parametrization, the electric current is carried by [using Eq. \eqref{currentdef} and  Eqs. (\ref{def1},\ref{def2})]
\begin{align}
\label{jsdef}
\mathbf{J}_s &= \sigma\nabla w-w\nabla\sigma+\frac{\sigma^2+w^2}{2\bar\rho_r}\nabla\theta_r,\\
\mathbf{J}_r &= \frac{(\bar\rho_r + \delta\rho_r)^2}{\bar\rho_r}\nabla\theta_r.
\end{align}
Defining Fourier transforms as in Eq.~\eqref{fourierdef} for $\sigma$, $w$,
$\delta\rho_r$ and $\theta_r$, we can write the effective action in Fourier
domain. From Eq.~\eqref{sc}, the action for the $\sigma$-field, undergoing the Ising transition, is given by
\begin{align}
S[\sigma] &= \sum_{\mathbf{q}} (\mathbf{q}^2+t_c)\sigma(\mathbf{q})\sigma(-\mathbf{q})+\frac{1}{2\Omega}\sum_{\mathbf{p}} u_c\nonumber\\&\qquad\sum_{\mathbf{q},\mathbf{q}'}\sigma(\mathbf{q})\sigma(\mathbf{q}')\sigma(\mathbf{p}-\mathbf{q}')\sigma(-\mathbf{p}-\mathbf{q}').
\end{align}
Since the fields $w,\delta\rho_r, \theta$ remain regular, we only need the action governing them to quadratic order. The action for the $w$-field is given by
\begin{equation}
S[w] = \sum_{\mathbf{q}}(\mathbf{q}^2+r_w)w(\mathbf{q})w(-\mathbf{q}),
\end{equation}
The mean-field form of $r_w$ is given by\cite{Roy2017}
\begin{align}
\label{rwdef}
r_w &= r_M+\alpha\sqrt{\frac{-2r_J}{u_r}}-\frac{\beta r_J}{2u_r}\\&=\alpha\sqrt{\frac{-8r_J}{u_r}},
\end{align}
where the last line is valid only on the Ising-transition line. The actions for $\delta\rho_r$ and $\theta_r$ are given by the Gaussian actions: 
\begin{eqnarray}
S[\delta\rho_r] &=& \sum_{\mathbf{q}}[q_z^2 + \tilde{K}_r(q_x^2+q_y^2) +r_{\delta\rho_r}]\delta\rho_r(\mathbf{q})\delta\rho_r(-\mathbf{q}),\nonumber\\
S[\theta_r] &=& \sum_{\mathbf{q}}[q_z^2 + \tilde{K}_r(q_x^2+q_y^2)]\theta_r(\mathbf{q})\theta_r(-\mathbf{q}),
\end{eqnarray}
where $r_{\delta\rho_r}=-2r_J>0$ in terms of the bare action parameters.  

To compute the conductivity, we again resort to a perturbation based on the 1/N-expansion. Much of the earlier formalism can be borrowed over to the real fields.\cite{Coleman1988, ZinnJustin2002} We consider an $N$-component $\sigma_\alpha$ field. Doing again a Hubbard-Stratonovich decoupling and resumming the bubble-diagrams, we get the effective free-field propagator for the $\sigma_\alpha$ field:
\begin{equation}
G^\sigma_0(q) =\frac{1}{2(\mathbf{q}^2 + \tilde{t}_c)},
\end{equation}
where $\tilde{t}_c$ is the renormalized gap-parameter that goes to zero at the phase-transition. The effective interaction is given by
\begin{equation}
\tilde{u}_c(\mathbf{k}) = \frac{u_c}{1+u_cN\Pi^\sigma(\mathbf{k})},
\end{equation}
where now, 
\begin{equation}
\Pi^\sigma(\mathbf{k}) = \int\frac{d^3q}{(2\pi)^3}G^\sigma_0(\mathbf{k}+\mathbf{q})G^\sigma_0(\mathbf{q}).
\end{equation}
To perform the 1/N computation, we replace $t_c, u_c$ by $\tilde{t}_c,
\tilde{u}_c(\mathbf{k})$ and evaluate Eq.~\eqref{sigmadef}. We drop the flavor indices from now on. As will be shown below, the superconducting density changes as the system undergoes the Ising phase transition from $\{2\pi\}$ to the $\{4\pi\}$ phases. This change depends on the gap parameter of the $w$-field. Hence, it is non-universal and depends on the superconducting density before the transition. Thus, we restrict ourselves only to the lowest non-vanishing order in 1/N expansion. 

To the lowest (zeroth) order in 1/N, we set $\tilde{u}_c(\mathbf{k})$ to zero
resulting in a Gaussian theory for all the fields. First, we calculate the
contribution to the conductivity from $\rho_s(k_z)$. In the current
parametrization, from Eq.~\eqref{rhodef}, one gets
\begin{align}
\rho^{(0)}_s(k_z) &= \frac{2}{\Omega}\sum_{\mathbf{q}}\{\langle\sigma(\mathbf{q})\sigma(-\mathbf{q})\rangle+\langle w(\mathbf{q})w(-\mathbf{q})\rangle\}\nonumber\\&-\frac{8}{\Omega}\sum_{\mathbf{p},\mathbf{q},\mathbf{q}'}\langle\sigma(\mathbf{p})\sigma(\mathbf{q})w(-\mathbf{k}-\mathbf{p})w(\mathbf{q}')\rangle p_x(q_x'-q_x),\nonumber
\end{align}
where we have used the definition of $\mathbf{J}_s$ [cf.\ Eq.~\eqref{jsdef}],
the fact that $\langle\theta_r(\mathbf{q})\rangle=0$ and neglected terms of
the order $1/\bar\rho_r^2$. By making the last approximation, we restrict
ourselves to the regime where the phase $\{2\pi\}$ is sufficiently
well-developed (recall that in terms of mean-field estimates, $\bar\rho_r =
\sqrt{-r_J/2u_r}$). Using the expressions for the propagators for $\sigma, w$,
we get
\begin{align}
\label{rho0}
\rho^{(0)}_s(k_z) &= \int\frac{d^3q}{(2\pi)^3}\Big[\frac{1}{\tilde{t}_c+\mathbf{q}^2}+\frac{1}{r_w+\mathbf{q}^2}\nonumber\\&\qquad-\frac{4q_x^2}{(\tilde{t}_c+\mathbf{q}^2)\{r_w+(\mathbf{k}+\mathbf{q})^2\}}\Big]
\end{align}
Using the result \cite{Cha1991} that 
\begin{equation}
\int\frac{d^3q}{(2\pi)^3}\frac{\partial}{\partial q_x}[q_xG^\sigma_0(q)]=0,
\end{equation}
one can rewrite Eq.~\eqref{rho0} at the critical point as
\footnote{We note that Eq.~\eqref{eq:rho_crit} is qualitatively different from
	the corresponding expression obtained for the  3D-XY transition. In the
	present case, the complex order parameter is given by	$\sigma+i w$, where
	only $\sigma$ undergoes a phase transition due to the finite gap $r_w$. This
	difference manifests itself in the fact that for finite $r_w$, $\rho_s^{(0)}$
	approaches a constant for $k_z\to0$ (corresponding to a superconducting
	response). On the other hand, for the XY-transition (with $r_w=0$) the
	leading term is proportional to $k_z$ (corresponding to a dissipative
	response).
}
\begin{align}\label{eq:rho_crit}
[\rho^{(0)}_s(k_z)]_{\rm{crit}} &= \int\frac{d^3q}{(2\pi)^3}\frac{2q_x^2}{\mathbf{q}^2}\Big[\frac{1}{\mathbf{q}^2}+\frac{1}{r_w+\mathbf{q}^2}\nonumber\\&\qquad-\frac{2}{r_w+(\mathbf{q}+\mathbf{k})^2}\Big].
\end{align}
The above integral can be evaluated exactly. After some algebra, we get
\begin{align}
\frac{[\rho^{0}_s(ik_z\rightarrow \omega+i0^+)]_{\rm{crit}}}{k_z} = \frac{\sqrt{r_w}}{6\pi}\frac{i}{\omega+i0^+}.
\end{align}
Upon taking the real part of this above equation, we find that this gives rise to a superconducting response with superconducting density $\delta n_s = \sqrt{r_w}/6\pi$. 

Next, we compute the contributions to conductivity from $\rho_{sr}$ and $\rho_r$. Since $\mathbf{k}=(0,0,k_z)$, it follows that
\begin{equation}
\int d^3 e^{i\mathbf{k}\cdot\mathbf{x}}\langle\partial_x\theta_r(\mathbf{x})\partial_x\theta_r(0)\rangle=0. 
\end{equation}
This implies $\rho_{sr}=0$. Finally, the contribution to conductivity from $\rho_r(k_z)$ is the usual superconducting response. It can be computed analogously and to leading order (deep in the $\{2\pi\}$ phase) is given by
\begin{equation}
\rho^{(0)}_r(k_z) = 2\bar\rho_r^2.
\end{equation}
Thus, the total contribution to the real part of the conductivity is given by
\begin{equation}
\label{sigmaval}
{\rm Re}[\sigma_{xx}] = \frac{\pi e^2}{\hbar}(\bar{n}_s+\delta n_s)\delta(\omega),
\end{equation}
where the first contribution $\bar{n}_s=8\bar\rho^2_r$ comes from the superconducting
density due to the usual superconductivity. The second denoted by $\delta
n_s=\sqrt{r_w}/6\pi$ is the renormalization of the superconducting density due to
the Ising transition. An estimate of the gap parameter for the $w$ field is
obtained by replacing $\psi_r$ by $\bar\rho_r$ in the action of
Eq.~\eqref{action_tot} with the parametrization of Eqs.~\eqref{def1} and \eqref{def2}. Then, $r_w = 2\alpha\bar\rho_r$. Thus, the fractional change in the superconducting density is given by:
\begin{align}
\label{nschange}
\frac{\delta n_s}{\bar{n}_s} = \frac{2}{3\pi}\Big(\frac{\alpha}{{\bar{n}_s}^3}\Big)^{1/2}.
\end{align}
The above equation relates the change in the superconducting density with the
coupling constant $\alpha$ and provides the following two important insights. Firstly,
the cubic coupling can be measured by measuring the change in the
superconducting density as one tunes through the Ising phase transition. A larger
jump in the density indicates a larger cubic coupling and vice versa.
Secondly, it also provides an estimate of the change in the superconducting density
as the Josephson coupling changes. It follows from Eq.~\eqref{nschange} that
as one increases the Josephson coupling, the change in the superconducting density decreases. This is because $\bar{n}_s$ increases \footnote{At some point in the parameter space, $\bar{n}_s$ saturates, but it is outside the validity of our field theory analysis.}. 

\subsection{Conductivities at the tricritical points}
\label{tps}
We concluding this section by briefly addressing the conductivities at the Ising and XY tricritical points. The upper critical dimension for the sextic interaction is three. Therefore, modulo logarithmic corrections because of the interaction being marginal, the universal quantities are given by the effective Gaussian theories. Thus, the results from the previous sections can be directly applied. 

For the 3D-XY tricritical point separating phases $\{\theta\}$ and $\{4\pi\}$ ($\rm{TP_1}$ in Fig. \ref{pd}), the conductivity can be obtained by
setting $N\rightarrow\infty$ (see Ref.~\onlinecite{Cha1991} for details). Therefore, the conductivity at this point is given by  
\begin{equation}
\sigma_{xx}=\frac{\pi}{8}\frac{e^2}{h}.
\end{equation}
Thus, the conductivity increases along the 3D-XY line and approaches the above value from the one given by Eq.~\eqref{sigmaval1} as one approaches the tricritical point. Similarly, the conductivity at the tricritical point separating phases $\{\theta\}$ and $\{2\pi\}$ ($\rm{TP_3}$ in Fig. \ref{pd}), the conductivity is given by 
\begin{equation}
\sigma_{xx}=\frac{\pi}{2}\frac{e^2}{h}.
\end{equation} 
 
At the 3D-Ising tricritical point ($\rm{TP_2}$ in Fig. \ref{pd}), the conductivity is given by Eq.~\eqref{sigmaval}. As discussed earlier, there is no dissipative component to the conductivity. Only the superconducting density changes across the transition. The change is largest at the tricritical point and decreases along the Ising phase-transition line. 

\section{Conclusion}
\label{concl}
To summarize, we have presented in this paper the charge-response of the Majorana toric code. From the basic microscopic Hamiltonian, after a Jordan-Wigner transformation, we mapped the problem to that of spins coupled to rotors. We computed a correction to the previously proposed field theory to one-loop order. Our calculations show that the phase-diagram is likely to contain another tricritical point of the 3D-XY type. Subsequently, we computed the conductivities at the different continuous phase-transitions using 1/N expansion. We provided the universal conductivities at the two 3D-XY phase-transitions and at the 3D-XY tricritical point. Next, we calculated the change in the superconducting density as the system undergoes the 3D-Ising phase-transition from a charge-$2e$ to a charge-$e$ superconductor. We showed that the change in the superconducting density provides an estimate of the nonlinear coupling between the spin and rotor fields. We emphasize that the calculated conductivities provide unique signatures of the different phase-transitions in the model. At the two 3D-XY transitions, the system behaves as a metal, where the conductivities have universal values. On the other hand, for the Ising transition, out of two coupled real fields responsible for carrying the current, only one undergoes a phase-transition. This results in a jump of the superconducting density across the transition. With the recent developments in detecting Majorana bound-states in solid-state systems, \cite{Mourik2012, Albrecht2016} we are optimistic of experimental verifications of our field theory predictions. 

Discussions with Barbara Terhal and David DiVincenzo are gratefully acknowledged. AR acknowledges the support through the ERC Consolidator Grant No. 682726 and the Alexander von Humboldt foundation. FH acknowledges the support of the Excellent Initiative of the Deutsche Forschungsgemeinschaft.

\appendix
\section{Mean-field analysis}
\label{mft}
In this section, we perform a mean-field calculation that provide supporting evidence for the proposed field theory of Ref. \onlinecite{Roy2017}. This can be done by first writing a Hubbard-Stratonovich decoupling in path-integral form and solving the corresponding Schr\"odinger problem\cite{Herbut2007, Sachdev2011} using perturbation theory. To that end, we introduce two complex order parameters, $\psi_s$ and $\psi_r$, respectively for the spin and rotor sectors of our model. Thus, at each lattice site, we define: 
\begin{align}
\sigma_i^+ &= \psi_s + (\sigma_i^+-\psi_s), \\ e^{i\phi_i} &= \psi_r + (e^{i\phi_i}-\psi_r).
\end{align}
We perform perturbation analysis with the unperturbed Hamiltonian $H_C$. Expanding around these order parameters and keeping to leading order, the mean field perturbation Hamiltonian is: 
\begin{align}
H_M^{\rm{mf}} + H_J^{\rm{mf}} &= -zE_M \sum_i \Big\{\big[2\psi_s^*\psi_r + \psi_s(1+|\psi_r|^2)\big]\sigma_i^-\nonumber\\&\quad+\big(\psi_s^2 + |\psi_s|^2\psi_r\big)e^{-i\phi_i} + {\rm{H.c.}}\Big\}\nonumber\\&\quad-zE_J \sum_i \big(\psi_r e^{-i\phi_i} + {\rm{H.c.}}\big),
\end{align}
where $z=4$ is the number of nearest neighbors around each lattice point. Since the order parameters are independent of the site index, the problem effectively reduces to that of a single site. Hence, we drop the site index. The ground state energy per site of the total interacting Hamiltonian can be written as: 
\begin{align}
\frac{E_g}{N_{\rm{tot}}} &= \frac{E_{\rm{mf}}}{N_{\rm{tot}}} - zE_M \big[\langle \sigma^-\rangle^2\langle e^{i\phi}\rangle + \langle \sigma^+\rangle^2\langle e^{-i\phi}\rangle\nonumber\\&+\langle \sigma^+\rangle\langle \sigma^-\rangle(1+\langle e^{i\phi}\rangle\langle e^{-i\phi}\rangle)\big]-zE_J \langle e^{i\phi}\rangle\langle e^{-i\phi}\rangle\nonumber\\&+zE_C\big[v_1\langle\sigma^-\rangle+v_1^*\langle\sigma^+\rangle+v_2\langle e^{-i\phi}\rangle +v_2^*\langle e^{i\phi}\rangle\big].\nonumber
\end{align}
Here $E_g$ is the total ground state energy, $N_{\rm{tot}}$ is the number of lattice sites, $E_{\rm{mf}}$ is the energy of the mean field model calculated using perturbation theory and 
\begin{align}
v_1 &= \frac{E_M}{E_C}\big[2\psi_s^*\psi_r + \psi_s(1+|\psi_r|^2)\big],\nonumber\\ v_2 &= \frac{E_M}{E_C}\big(\psi_s^2 + |\psi_s|^2\psi_r\big)+\frac{E_J}{E_C}\psi_r.
\end{align}
We perform this computation till fourth order in order to get the relevant
terms of the field theory given in Eq.~\eqref{action_tot}. The mean field energy $E_{\rm{mf}}$ in units of $E_C$ is
\begin{align}
\label{energy}
\frac{E_{\rm{mf}}}{N_{\rm{tot}}E_C} &= -z^2|v_1|^2-\frac{z^2|v_2|^2}{2}+z^4|v_1|^4+\frac{7z^4|v_2|^4}{128}\nonumber\\&\quad-\frac{10z^4|v_1v_2|^2}{9},
\end{align}
The ground state is computed to third order in perturbation theory, the explicit form of which is not shown for brevity. Collecting all the terms, one has for the ground state energy:
\begin{align}
\frac{E_g}{N_{\rm{tot}}E_C}&=16\Big(1-\frac{4E_M}{E_C}\Big)|v_1|^2+8\Big(1-\frac{2E_J}{E_C}\Big)|v_2|^2\nonumber\\&\quad+256|v_1|^4+\Big(6+\frac{16E_J}{E_C}\Big)|v_2|^4\nonumber\\
&\quad
+\frac{128}{9}\Big(-145+\frac{152E_J}{E_C}+\frac{178E_M}{E_C}\Big)|v_1v_2|^2\nonumber\\&
\quad-\frac{128E_M}{E_C}(v_1^2v_2+{v_1^*}^2v_2^*),
\end{align}
where we have set $z=4$. In terms of the order parameters $\psi_s, \psi_r$, to fourth order, we get
\begin{align}
\frac{E_g}{N_{\rm{tot}}E_C}&=\Big(1-\frac{4E_M}{E_C}\Big)|\psi_s|^2+2\Big(1-\frac{2E_J}{E_C}\Big)|\psi_r|^2\nonumber\\&\quad +2|\psi_s|^4+\frac{3}{8}|\psi_r|^4+18|\psi_s|^2|\psi_r|^2\nonumber\\&\quad+3(\psi_s^{*2}\psi_r + \psi_s^2\psi_r^*).
\end{align}
In the coefficients of the higher than quadratic terms, we have kept only the
leading order contributions. The dependence of the ground state energy on the
order parameters further justifies the choice of the form of the action argued
in Ref.~\onlinecite{Roy2017} purely on the basis of symmetries. The mean-field estimates of the action parameters are
\begin{align}
r_M &= 1-\frac{4E_M}{E_C},\ r_J = 2\Big(1-\frac{2E_J}{E_C}\Big), \nonumber\\
u_s &= 2, \ u_r = \frac{3}{8},\ |\alpha| = 3,\ \beta = 18,
\end{align}
where we have set the lattice-constant to unity. Therefore, the mean-field estimates of the location of the phase-transition between the $\{\theta\}$ and the $\{4\pi\}$ phases is given by at $E_M = E_C/4$. The same for the phase-transition between the $\{\theta\}$ and the $\{2\pi\}$ phases is $E_J = E_C/2$.

\bibliography{library_1}

\begin{thebibliography}{48}%
\makeatletter
\providecommand \@ifxundefined [1]{%
 \@ifx{#1\undefined}
}%
\providecommand \@ifnum [1]{%
 \ifnum #1\expandafter \@firstoftwo
 \else \expandafter \@secondoftwo
 \fi
}%
\providecommand \@ifx [1]{%
 \ifx #1\expandafter \@firstoftwo
 \else \expandafter \@secondoftwo
 \fi
}%
\providecommand \natexlab [1]{#1}%
\providecommand \enquote  [1]{``#1''}%
\providecommand \bibnamefont  [1]{#1}%
\providecommand \bibfnamefont [1]{#1}%
\providecommand \citenamefont [1]{#1}%
\providecommand \href@noop [0]{\@secondoftwo}%
\providecommand \href [0]{\begingroup \@sanitize@url \@href}%
\providecommand \@href[1]{\@@startlink{#1}\@@href}%
\providecommand \@@href[1]{\endgroup#1\@@endlink}%
\providecommand \@sanitize@url [0]{\catcode `\\12\catcode `\$12\catcode
  `\&12\catcode `\#12\catcode `\^12\catcode `\_12\catcode `\%12\relax}%
\providecommand \@@startlink[1]{}%
\providecommand \@@endlink[0]{}%
\providecommand \url  [0]{\begingroup\@sanitize@url \@url }%
\providecommand \@url [1]{\endgroup\@href {#1}{\urlprefix }}%
\providecommand \urlprefix  [0]{URL }%
\providecommand \Eprint [0]{\href }%
\providecommand \doibase [0]{http://dx.doi.org/}%
\providecommand \selectlanguage [0]{\@gobble}%
\providecommand \bibinfo  [0]{\@secondoftwo}%
\providecommand \bibfield  [0]{\@secondoftwo}%
\providecommand \translation [1]{[#1]}%
\providecommand \BibitemOpen [0]{}%
\providecommand \bibitemStop [0]{}%
\providecommand \bibitemNoStop [0]{.\EOS\space}%
\providecommand \EOS [0]{\spacefactor3000\relax}%
\providecommand \BibitemShut  [1]{\csname bibitem#1\endcsname}%
\let\auto@bib@innerbib\@empty
\bibitem [{\citenamefont {DiVincenzo}(2009)}]{DiVincenzo2009}%
  \BibitemOpen
  \bibfield  {author} {\bibinfo {author} {\bibfnamefont {D.~P.}\ \bibnamefont
  {DiVincenzo}},\ }\href {\doibase 10.1088/0031-8949/2009/T137/014020}
  {\bibfield  {journal} {\bibinfo  {journal} {Phys. Scripta}\ }\textbf
  {\bibinfo {volume} {T137}},\ \bibinfo {pages} {014020} (\bibinfo {year}
  {2009})}\BibitemShut {NoStop}%
\bibitem [{\citenamefont {Kitaev}(2003)}]{Kitaev2003}%
  \BibitemOpen
  \bibfield  {author} {\bibinfo {author} {\bibfnamefont {A.}~\bibnamefont
  {Kitaev}},\ }\href {\doibase 10.1016/S0003-4916(02)00018-0} {\bibfield
  {journal} {\bibinfo  {journal} {Ann. Phys. (NY)}\ }\textbf {\bibinfo {volume}
  {303}},\ \bibinfo {pages} {2} (\bibinfo {year} {2003})}\BibitemShut {NoStop}%
\bibitem [{\citenamefont {Kitaev}(2006)}]{Kitaev2006}%
  \BibitemOpen
  \bibfield  {author} {\bibinfo {author} {\bibfnamefont {A.}~\bibnamefont
  {Kitaev}},\ }\href {\doibase 10.1016/j.aop.2005.10.005} {\bibfield  {journal}
  {\bibinfo  {journal} {Ann. Phys. (NY)}\ }\textbf {\bibinfo {volume} {321}},\
  \bibinfo {pages} {2} (\bibinfo {year} {2006})}\BibitemShut {NoStop}%
\bibitem [{\citenamefont {Fowler}\ \emph {et~al.}(2012)\citenamefont {Fowler},
  \citenamefont {Mariantoni}, \citenamefont {Martinis},\ and\ \citenamefont
  {Cleland}}]{Fowler_Cleland_2012}%
  \BibitemOpen
  \bibfield  {author} {\bibinfo {author} {\bibfnamefont {A.~G.}\ \bibnamefont
  {Fowler}}, \bibinfo {author} {\bibfnamefont {M.}~\bibnamefont {Mariantoni}},
  \bibinfo {author} {\bibfnamefont {J.~M.}\ \bibnamefont {Martinis}}, \ and\
  \bibinfo {author} {\bibfnamefont {A.~N.}\ \bibnamefont {Cleland}},\ }\href
  {\doibase 10.1103/PhysRevA.86.032324} {\bibfield  {journal} {\bibinfo
  {journal} {Phys. Rev. A}\ }\textbf {\bibinfo {volume} {86}} (\bibinfo {year}
  {2012}),\ 10.1103/PhysRevA.86.032324}\BibitemShut {NoStop}%
\bibitem [{\citenamefont {Levin}\ and\ \citenamefont {Wen}(2006)}]{Levin2006}%
  \BibitemOpen
  \bibfield  {author} {\bibinfo {author} {\bibfnamefont {M.}~\bibnamefont
  {Levin}}\ and\ \bibinfo {author} {\bibfnamefont {X.-G.}\ \bibnamefont
  {Wen}},\ }\href {\doibase 10.1103/PhysRevLett.96.110405} {\bibfield
  {journal} {\bibinfo  {journal} {Phys. Rev. Lett.}\ }\textbf {\bibinfo
  {volume} {96}},\ \bibinfo {pages} {110405} (\bibinfo {year}
  {2006})}\BibitemShut {NoStop}%
\bibitem [{\citenamefont {Xu}\ and\ \citenamefont {Fu}(2010)}]{Xu2010}%
  \BibitemOpen
  \bibfield  {author} {\bibinfo {author} {\bibfnamefont {C.}~\bibnamefont
  {Xu}}\ and\ \bibinfo {author} {\bibfnamefont {L.}~\bibnamefont {Fu}},\ }\href
  {\doibase 10.1103/PhysRevB.81.134435} {\bibfield  {journal} {\bibinfo
  {journal} {Phys. Rev. B}\ }\textbf {\bibinfo {volume} {81}},\ \bibinfo
  {pages} {1} (\bibinfo {year} {2010})}\BibitemShut {NoStop}%
\bibitem [{\citenamefont {Terhal}\ \emph {et~al.}(2012)\citenamefont {Terhal},
  \citenamefont {Hassler},\ and\ \citenamefont {Divincenzo}}]{Terhal2012}%
  \BibitemOpen
  \bibfield  {author} {\bibinfo {author} {\bibfnamefont {B.~M.}\ \bibnamefont
  {Terhal}}, \bibinfo {author} {\bibfnamefont {F.}~\bibnamefont {Hassler}}, \
  and\ \bibinfo {author} {\bibfnamefont {D.~P.}\ \bibnamefont {Divincenzo}},\
  }\href {\doibase 10.1103/PhysRevLett.108.260504} {\bibfield  {journal}
  {\bibinfo  {journal} {Phys. Rev. Lett.}\ }\textbf {\bibinfo {volume} {108}},\
  \bibinfo {pages} {1} (\bibinfo {year} {2012})}\BibitemShut {NoStop}%
\bibitem [{\citenamefont {Nussinov}\ \emph {et~al.}(2012)\citenamefont
  {Nussinov}, \citenamefont {Ortiz},\ and\ \citenamefont
  {Cobanera}}]{Nussinov2012}%
  \BibitemOpen
  \bibfield  {author} {\bibinfo {author} {\bibfnamefont {Z.}~\bibnamefont
  {Nussinov}}, \bibinfo {author} {\bibfnamefont {G.}~\bibnamefont {Ortiz}}, \
  and\ \bibinfo {author} {\bibfnamefont {E.}~\bibnamefont {Cobanera}},\ }\href
  {\doibase 10.1103/PhysRevB.86.085415} {\bibfield  {journal} {\bibinfo
  {journal} {Phys. Rev. B}\ }\textbf {\bibinfo {volume} {86}},\ \bibinfo
  {pages} {085415} (\bibinfo {year} {2012})}\BibitemShut {NoStop}%
\bibitem [{\citenamefont {Vijay}\ \emph {et~al.}(2015)\citenamefont {Vijay},
  \citenamefont {Hsieh},\ and\ \citenamefont {Fu}}]{Vijay2015}%
  \BibitemOpen
  \bibfield  {author} {\bibinfo {author} {\bibfnamefont {S.}~\bibnamefont
  {Vijay}}, \bibinfo {author} {\bibfnamefont {T.~H.}\ \bibnamefont {Hsieh}}, \
  and\ \bibinfo {author} {\bibfnamefont {L.}~\bibnamefont {Fu}},\ }\href
  {\doibase 10.1103/PhysRevX.5.041038} {\bibfield  {journal} {\bibinfo
  {journal} {Phys. Rev. X}\ }\textbf {\bibinfo {volume} {5}},\ \bibinfo {pages}
  {41038} (\bibinfo {year} {2015})}\BibitemShut {NoStop}%
\bibitem [{\citenamefont {Landau}\ \emph {et~al.}(2016)\citenamefont {Landau},
  \citenamefont {Plugge}, \citenamefont {Sela}, \citenamefont {Altland},
  \citenamefont {Albrecht},\ and\ \citenamefont {Egger}}]{Landau2016}%
  \BibitemOpen
  \bibfield  {author} {\bibinfo {author} {\bibfnamefont {L.~A.}\ \bibnamefont
  {Landau}}, \bibinfo {author} {\bibfnamefont {S.}~\bibnamefont {Plugge}},
  \bibinfo {author} {\bibfnamefont {E.}~\bibnamefont {Sela}}, \bibinfo {author}
  {\bibfnamefont {A.}~\bibnamefont {Altland}}, \bibinfo {author} {\bibfnamefont
  {S.~M.}\ \bibnamefont {Albrecht}}, \ and\ \bibinfo {author} {\bibfnamefont
  {R.}~\bibnamefont {Egger}},\ }\href {\doibase 10.1103/PhysRevLett.116.050501}
  {\bibfield  {journal} {\bibinfo  {journal} {Phys. Rev. Lett.}\ }\textbf
  {\bibinfo {volume} {116}},\ \bibinfo {pages} {1} (\bibinfo {year}
  {2016})}\BibitemShut {NoStop}%
\bibitem [{\citenamefont {Karzig}\ \emph {et~al.}(2017)\citenamefont {Karzig},
  \citenamefont {Knapp}, \citenamefont {Lutchyn}, \citenamefont {Bonderson},
  \citenamefont {Hastings}, \citenamefont {Nayak}, \citenamefont {Alicea},
  \citenamefont {Flensberg}, \citenamefont {Plugge}, \citenamefont {Oreg},
  \citenamefont {Marcus},\ and\ \citenamefont {Freedman}}]{Karzig2016}%
  \BibitemOpen
  \bibfield  {author} {\bibinfo {author} {\bibfnamefont {T.}~\bibnamefont
  {Karzig}}, \bibinfo {author} {\bibfnamefont {C.}~\bibnamefont {Knapp}},
  \bibinfo {author} {\bibfnamefont {R.~M.}\ \bibnamefont {Lutchyn}}, \bibinfo
  {author} {\bibfnamefont {P.}~\bibnamefont {Bonderson}}, \bibinfo {author}
  {\bibfnamefont {M.~B.}\ \bibnamefont {Hastings}}, \bibinfo {author}
  {\bibfnamefont {C.}~\bibnamefont {Nayak}}, \bibinfo {author} {\bibfnamefont
  {J.}~\bibnamefont {Alicea}}, \bibinfo {author} {\bibfnamefont
  {K.}~\bibnamefont {Flensberg}}, \bibinfo {author} {\bibfnamefont
  {S.}~\bibnamefont {Plugge}}, \bibinfo {author} {\bibfnamefont
  {Y.}~\bibnamefont {Oreg}}, \bibinfo {author} {\bibfnamefont {C.~M.}\
  \bibnamefont {Marcus}}, \ and\ \bibinfo {author} {\bibfnamefont {M.~H.}\
  \bibnamefont {Freedman}},\ }\href {\doibase 10.1103/PhysRevB.95.235305}
  {\bibfield  {journal} {\bibinfo  {journal} {Phys. Rev. B}\ }\textbf {\bibinfo
  {volume} {95}},\ \bibinfo {pages} {235305} (\bibinfo {year}
  {2017})}\BibitemShut {NoStop}%
\bibitem [{\citenamefont {Litinski}\ \emph {et~al.}(2017)\citenamefont
  {Litinski}, \citenamefont {Kesselring}, \citenamefont {Eisert},\ and\
  \citenamefont {{Von Oppen}}}]{Litinski2017}%
  \BibitemOpen
  \bibfield  {author} {\bibinfo {author} {\bibfnamefont {D.}~\bibnamefont
  {Litinski}}, \bibinfo {author} {\bibfnamefont {M.~S.}\ \bibnamefont
  {Kesselring}}, \bibinfo {author} {\bibfnamefont {J.}~\bibnamefont {Eisert}},
  \ and\ \bibinfo {author} {\bibfnamefont {F.}~\bibnamefont {{Von Oppen}}},\
  }\href {https://arxiv.org/pdf/1704.01589.pdf} {\bibfield  {journal} {\bibinfo
   {journal} {arXiv:1704.01589}\ } (\bibinfo {year} {2017})}\BibitemShut
  {NoStop}%
\bibitem [{\citenamefont {Kitaev}(2001)}]{Kitaev2000}%
  \BibitemOpen
  \bibfield  {author} {\bibinfo {author} {\bibfnamefont {A.~{\relax Yu}.}\
  \bibnamefont {Kitaev}},\ }\href@noop {} {\bibfield  {journal} {\bibinfo
  {journal} {Phys.-Usp.}\ }\textbf {\bibinfo {volume} {44}},\ \bibinfo {pages}
  {131} (\bibinfo {year} {2001})}\BibitemShut {NoStop}%
\bibitem [{\citenamefont {Fu}(2010)}]{Fu2010}%
  \BibitemOpen
  \bibfield  {author} {\bibinfo {author} {\bibfnamefont {L.}~\bibnamefont
  {Fu}},\ }\href {\doibase 10.1103/PhysRevLett.104.056402} {\bibfield
  {journal} {\bibinfo  {journal} {Phys. Rev. Lett.}\ }\textbf {\bibinfo
  {volume} {104}},\ \bibinfo {pages} {1} (\bibinfo {year} {2010})}\BibitemShut
  {NoStop}%
\bibitem [{\citenamefont {Roy}\ \emph {et~al.}(2017)\citenamefont {Roy},
  \citenamefont {Terhal},\ and\ \citenamefont {Hassler}}]{Roy2017}%
  \BibitemOpen
  \bibfield  {author} {\bibinfo {author} {\bibfnamefont {A.}~\bibnamefont
  {Roy}}, \bibinfo {author} {\bibfnamefont {B.~M.}\ \bibnamefont {Terhal}}, \
  and\ \bibinfo {author} {\bibfnamefont {F.}~\bibnamefont {Hassler}},\ }\href
  {\doibase 10.1103/PhysRevLett.119.180508} {\bibfield  {journal} {\bibinfo
  {journal} {Phys. Rev. Lett.}\ }\textbf {\bibinfo {volume} {119}},\ \bibinfo
  {pages} {180508} (\bibinfo {year} {2017})}\BibitemShut {NoStop}%
\bibitem [{\citenamefont {Fisher}\ \emph {et~al.}(1989)\citenamefont {Fisher},
  \citenamefont {Weichman}, \citenamefont {Grinstein},\ and\ \citenamefont
  {Fisher}}]{Fisher1989}%
  \BibitemOpen
  \bibfield  {author} {\bibinfo {author} {\bibfnamefont {M.~P.~A.}\
  \bibnamefont {Fisher}}, \bibinfo {author} {\bibfnamefont {P.~B.}\
  \bibnamefont {Weichman}}, \bibinfo {author} {\bibfnamefont {G.}~\bibnamefont
  {Grinstein}}, \ and\ \bibinfo {author} {\bibfnamefont {D.~S.}\ \bibnamefont
  {Fisher}},\ }\href {\doibase 10.1103/PhysRevB.40.546} {\bibfield  {journal}
  {\bibinfo  {journal} {Phys. Rev. B}\ }\textbf {\bibinfo {volume} {40}},\
  \bibinfo {pages} {546} (\bibinfo {year} {1989})}\BibitemShut {NoStop}%
\bibitem [{\citenamefont {Fazio}\ and\ \citenamefont
  {Sch{\"o}n}(1998)}]{Fazio1997}%
  \BibitemOpen
  \bibfield  {author} {\bibinfo {author} {\bibfnamefont {R.}~\bibnamefont
  {Fazio}}\ and\ \bibinfo {author} {\bibfnamefont {G.}~\bibnamefont
  {Sch{\"o}n}},\ }\href {\doibase 10.1063/1.55278} {\bibfield  {journal}
  {\bibinfo  {journal} {AIP Conf. Proc.}\ }\textbf {\bibinfo {volume} {427}},\
  \bibinfo {pages} {273} (\bibinfo {year} {1998})}\BibitemShut {NoStop}%
\bibitem [{\citenamefont {Fazio}\ and\ \citenamefont {van~der
  Zant}(2001)}]{Fazio2001}%
  \BibitemOpen
  \bibfield  {author} {\bibinfo {author} {\bibfnamefont {R.}~\bibnamefont
  {Fazio}}\ and\ \bibinfo {author} {\bibfnamefont {H.}~\bibnamefont {van~der
  Zant}},\ }\href {\doibase 10.1016/S0370-1573(01)00022-9} {\bibfield
  {journal} {\bibinfo  {journal} {Phys. Rep.}\ }\textbf {\bibinfo {volume}
  {355}},\ \bibinfo {pages} {235} (\bibinfo {year} {2001})}\BibitemShut
  {NoStop}%
\bibitem [{\citenamefont {Herbut}(2007)}]{Herbut2007}%
  \BibitemOpen
  \bibfield  {author} {\bibinfo {author} {\bibfnamefont {I.}~\bibnamefont
  {Herbut}},\ }\href@noop {} {\emph {\bibinfo {title} {{A Modern Approach to
  Critical Phenomena}}}}\ (\bibinfo  {publisher} {Cambridge University Press},\
  \bibinfo {year} {2007})\BibitemShut {NoStop}%
\bibitem [{\citenamefont {Sachdev}(2011)}]{Sachdev2011}%
  \BibitemOpen
  \bibfield  {author} {\bibinfo {author} {\bibfnamefont {S.}~\bibnamefont
  {Sachdev}},\ }\href {https://books.google.de/books?id=F3IkpxwpqSgC} {\emph
  {\bibinfo {title} {{Quantum Phase Transitions}}}}\ (\bibinfo  {publisher}
  {Cambridge University Press},\ \bibinfo {year} {2011})\BibitemShut {NoStop}%
\bibitem [{\citenamefont {Wen}(1991)}]{Wen1991}%
  \BibitemOpen
  \bibfield  {author} {\bibinfo {author} {\bibfnamefont {X.~G.}\ \bibnamefont
  {Wen}},\ }\href {\doibase 10.1103/PhysRevB.44.2664} {\bibfield  {journal}
  {\bibinfo  {journal} {Phys. Rev. B}\ }\textbf {\bibinfo {volume} {44}},\
  \bibinfo {pages} {2664} (\bibinfo {year} {1991})}\BibitemShut {NoStop}%
\bibitem [{\citenamefont {Balents}\ \emph {et~al.}(1999)\citenamefont
  {Balents}, \citenamefont {Fisher},\ and\ \citenamefont
  {Nayak}}]{Balents1999}%
  \BibitemOpen
  \bibfield  {author} {\bibinfo {author} {\bibfnamefont {L.}~\bibnamefont
  {Balents}}, \bibinfo {author} {\bibfnamefont {M.~P.~A.}\ \bibnamefont
  {Fisher}}, \ and\ \bibinfo {author} {\bibfnamefont {C.}~\bibnamefont
  {Nayak}},\ }\href {\doibase 10.1103/PhysRevB.60.1654} {\bibfield  {journal}
  {\bibinfo  {journal} {Phys. Rev. B}\ }\textbf {\bibinfo {volume} {60}},\
  \bibinfo {pages} {1654} (\bibinfo {year} {1999})}\BibitemShut {NoStop}%
\bibitem [{\citenamefont {Senthil}\ and\ \citenamefont
  {Fisher}(2000)}]{Senthil2000}%
  \BibitemOpen
  \bibfield  {author} {\bibinfo {author} {\bibfnamefont {T.}~\bibnamefont
  {Senthil}}\ and\ \bibinfo {author} {\bibfnamefont {M.~P.~A.}\ \bibnamefont
  {Fisher}},\ }\href {\doibase 10.1103/PhysRevB.62.7850} {\bibfield  {journal}
  {\bibinfo  {journal} {Phys. Rev. B}\ }\textbf {\bibinfo {volume} {62}},\
  \bibinfo {pages} {7850} (\bibinfo {year} {2000})}\BibitemShut {NoStop}%
\bibitem [{\citenamefont {Cha}\ \emph {et~al.}(1991)\citenamefont {Cha},
  \citenamefont {Fisher}, \citenamefont {Girvin}, \citenamefont {Wallin},\ and\
  \citenamefont {Young}}]{Cha1991}%
  \BibitemOpen
  \bibfield  {author} {\bibinfo {author} {\bibfnamefont {M.-C.}\ \bibnamefont
  {Cha}}, \bibinfo {author} {\bibfnamefont {M.~P.~A.}\ \bibnamefont {Fisher}},
  \bibinfo {author} {\bibfnamefont {S.~M.}\ \bibnamefont {Girvin}}, \bibinfo
  {author} {\bibfnamefont {M.}~\bibnamefont {Wallin}}, \ and\ \bibinfo {author}
  {\bibfnamefont {A.~P.}\ \bibnamefont {Young}},\ }\href {\doibase
  10.1103/PhysRevB.44.6883} {\bibfield  {journal} {\bibinfo  {journal} {Phys.
  Rev. B}\ }\textbf {\bibinfo {volume} {44}},\ \bibinfo {pages} {6883}
  (\bibinfo {year} {1991})}\BibitemShut {NoStop}%
\bibitem [{\citenamefont {Fazio}\ and\ \citenamefont
  {Zappal\`a}(1996)}]{Fazio1996}%
  \BibitemOpen
  \bibfield  {author} {\bibinfo {author} {\bibfnamefont {R.}~\bibnamefont
  {Fazio}}\ and\ \bibinfo {author} {\bibfnamefont {D.}~\bibnamefont
  {Zappal\`a}},\ }\href {\doibase 10.1103/PhysRevB.53.R8883} {\bibfield
  {journal} {\bibinfo  {journal} {Phys. Rev. B}\ }\textbf {\bibinfo {volume}
  {53}},\ \bibinfo {pages} {R8883} (\bibinfo {year} {1996})}\BibitemShut
  {NoStop}%
\bibitem [{\citenamefont {Chamon}\ \emph {et~al.}(2010)\citenamefont {Chamon},
  \citenamefont {Jackiw}, \citenamefont {Nishida}, \citenamefont {Pi},\ and\
  \citenamefont {Santos}}]{Chamon2010}%
  \BibitemOpen
  \bibfield  {author} {\bibinfo {author} {\bibfnamefont {C.}~\bibnamefont
  {Chamon}}, \bibinfo {author} {\bibfnamefont {R.}~\bibnamefont {Jackiw}},
  \bibinfo {author} {\bibfnamefont {Y.}~\bibnamefont {Nishida}}, \bibinfo
  {author} {\bibfnamefont {S.-Y.}\ \bibnamefont {Pi}}, \ and\ \bibinfo {author}
  {\bibfnamefont {L.}~\bibnamefont {Santos}},\ }\href {\doibase
  10.1103/PhysRevB.81.224515} {\bibfield  {journal} {\bibinfo  {journal} {Phys.
  Rev. B}\ }\textbf {\bibinfo {volume} {81}},\ \bibinfo {pages} {224515}
  (\bibinfo {year} {2010})}\BibitemShut {NoStop}%
\bibitem [{\citenamefont {Beenakker}(2014)}]{Beenakker2014}%
  \BibitemOpen
  \bibfield  {author} {\bibinfo {author} {\bibfnamefont {C.~W.~J.}\
  \bibnamefont {Beenakker}},\ }\href {\doibase 10.1103/PhysRevLett.112.070604}
  {\bibfield  {journal} {\bibinfo  {journal} {Phys.Rev.Lett.}\ }\textbf
  {\bibinfo {volume} {112}},\ \bibinfo {pages} {070604} (\bibinfo {year}
  {2014})}\BibitemShut {NoStop}%
\bibitem [{\citenamefont {Read}\ and\ \citenamefont {Green}(2000)}]{Read2000}%
  \BibitemOpen
  \bibfield  {author} {\bibinfo {author} {\bibfnamefont {N.}~\bibnamefont
  {Read}}\ and\ \bibinfo {author} {\bibfnamefont {D.}~\bibnamefont {Green}},\
  }\href@noop {} {\bibfield  {journal} {\bibinfo  {journal} {Phys. Rev. B}\
  }\textbf {\bibinfo {volume} {61}},\ \bibinfo {pages} {10267} (\bibinfo {year}
  {2000})}\BibitemShut {NoStop}%
\bibitem [{\citenamefont {Ivanov}(2001)}]{Ivanov2001}%
  \BibitemOpen
  \bibfield  {author} {\bibinfo {author} {\bibfnamefont {D.~A.}\ \bibnamefont
  {Ivanov}},\ }\href {\doibase 10.1103/PhysRevLett.86.268} {\bibfield
  {journal} {\bibinfo  {journal} {Phys. Rev. Lett.}\ }\textbf {\bibinfo
  {volume} {86}},\ \bibinfo {pages} {268} (\bibinfo {year} {2001})}\BibitemShut
  {NoStop}%
\bibitem [{\citenamefont {Nayak}\ \emph {et~al.}(2008)\citenamefont {Nayak},
  \citenamefont {Simon}, \citenamefont {Stern}, \citenamefont {Freedman},\ and\
  \citenamefont {{Das Sarma}}}]{Nayak2008}%
  \BibitemOpen
  \bibfield  {author} {\bibinfo {author} {\bibfnamefont {C.}~\bibnamefont
  {Nayak}}, \bibinfo {author} {\bibfnamefont {S.~H.}\ \bibnamefont {Simon}},
  \bibinfo {author} {\bibfnamefont {A.}~\bibnamefont {Stern}}, \bibinfo
  {author} {\bibfnamefont {M.}~\bibnamefont {Freedman}}, \ and\ \bibinfo
  {author} {\bibfnamefont {S.}~\bibnamefont {{Das Sarma}}},\ }\href {\doibase
  10.1103/RevModPhys.80.1083} {\bibfield  {journal} {\bibinfo  {journal}
  {Reviews of Modern Physics}\ }\textbf {\bibinfo {volume} {80}},\ \bibinfo
  {pages} {1083} (\bibinfo {year} {2008})}\BibitemShut {NoStop}%
\bibitem [{\citenamefont {Beenakker}(2013)}]{Beenakker2013}%
  \BibitemOpen
  \bibfield  {author} {\bibinfo {author} {\bibfnamefont {C.~W.~J.}\
  \bibnamefont {Beenakker}},\ }\href {\doibase
  10.1146/annurev-conmatphys-030212-184337} {\bibfield  {journal} {\bibinfo
  {journal} {Annual Review of Condensed Matter Physics}\ }\textbf {\bibinfo
  {volume} {4}},\ \bibinfo {pages} {15} (\bibinfo {year} {2013})}\BibitemShut
  {NoStop}%
\bibitem [{\citenamefont {van Heck}\ \emph {et~al.}(2011)\citenamefont {van
  Heck}, \citenamefont {Hassler}, \citenamefont {Akhmerov},\ and\ \citenamefont
  {Beenakker}}]{VanHeck2011}%
  \BibitemOpen
  \bibfield  {author} {\bibinfo {author} {\bibfnamefont {B.}~\bibnamefont {van
  Heck}}, \bibinfo {author} {\bibfnamefont {F.}~\bibnamefont {Hassler}},
  \bibinfo {author} {\bibfnamefont {A.~R.}\ \bibnamefont {Akhmerov}}, \ and\
  \bibinfo {author} {\bibfnamefont {C.~W.~J.}\ \bibnamefont {Beenakker}},\
  }\href {\doibase 10.1103/PhysRevB.84.180502} {\bibfield  {journal} {\bibinfo
  {journal} {Phys. Rev. B}\ }\textbf {\bibinfo {volume} {84}},\ \bibinfo
  {pages} {2} (\bibinfo {year} {2011})}\BibitemShut {NoStop}%
\bibitem [{\citenamefont {Alicea}(2012)}]{Alicea2012}%
  \BibitemOpen
  \bibfield  {author} {\bibinfo {author} {\bibfnamefont {J.}~\bibnamefont
  {Alicea}},\ }\href@noop {} {\bibfield  {journal} {\bibinfo  {journal} {Rep.
  Prog. Phys.}\ }\textbf {\bibinfo {volume} {75}},\ \bibinfo {pages} {076501}
  (\bibinfo {year} {2012})}\BibitemShut {NoStop}%
\bibitem [{Note1()}]{Note1}%
  \BibitemOpen
  \bibinfo {note} {Note that for case of $2n$ Majorana zero modes, the
  remaining topological degeneracy will be $2^{n-1}$.}\BibitemShut {Stop}%
\bibitem [{\citenamefont {Bravyi}(2006)}]{Bravyi2006}%
  \BibitemOpen
  \bibfield  {author} {\bibinfo {author} {\bibfnamefont {S.}~\bibnamefont
  {Bravyi}},\ }\href {\doibase 10.1103/PhysRevA.73.042313} {\bibfield
  {journal} {\bibinfo  {journal} {Phys. Rev. A}\ }\textbf {\bibinfo {volume}
  {73}},\ \bibinfo {pages} {1} (\bibinfo {year} {2006})}\BibitemShut {NoStop}%
\bibitem [{\citenamefont {van Heck}\ \emph {et~al.}(2012)\citenamefont {van
  Heck}, \citenamefont {Akhmerov}, \citenamefont {Hassler}, \citenamefont
  {Burrello},\ and\ \citenamefont {Beenakker}}]{VanHeck2012}%
  \BibitemOpen
  \bibfield  {author} {\bibinfo {author} {\bibfnamefont {B.}~\bibnamefont {van
  Heck}}, \bibinfo {author} {\bibfnamefont {A.~R.}\ \bibnamefont {Akhmerov}},
  \bibinfo {author} {\bibfnamefont {F.}~\bibnamefont {Hassler}}, \bibinfo
  {author} {\bibfnamefont {M.}~\bibnamefont {Burrello}}, \ and\ \bibinfo
  {author} {\bibfnamefont {C.~W.~J.}\ \bibnamefont {Beenakker}},\ }\href
  {\doibase 10.1088/1367-2630/14/3/035019} {\bibfield  {journal} {\bibinfo
  {journal} {New J. Phys.}\ }\textbf {\bibinfo {volume} {14}},\ \bibinfo
  {pages} {035019} (\bibinfo {year} {2012})}\BibitemShut {NoStop}%
\bibitem [{Coo()}]{Cooper_pair_note}%
  \BibitemOpen
  \href@noop {} {}\bibinfo {note} {Note that while $\phi_i$ and $n_i$ remain
  canonically conjugate, $n_i$ now denotes only the excess number of
  Cooper-pairs on the $i$-th island.}\BibitemShut {Stop}%
\bibitem [{Note2()}]{Note2}%
  \BibitemOpen
  \bibinfo {note} {Note that $\DOTSB \prod@ \slimits@ \DOTSB \prod@ \slimits@
  _\square s_{ij} = 1$ is trivially true.}\BibitemShut {Stop}%
\bibitem [{\citenamefont {Fradkin}(2013)}]{Fradkin2013}%
  \BibitemOpen
  \bibfield  {author} {\bibinfo {author} {\bibfnamefont {E.}~\bibnamefont
  {Fradkin}},\ }\href {https://books.google.de/books?id=x7\_6MX4ye\_wC} {\emph
  {\bibinfo {title} {Field Theories of Condensed Matter Physics}}},\ Field
  Theories of Condensed Matter Physics\ (\bibinfo  {publisher} {Cambridge
  University Press},\ \bibinfo {year} {2013})\BibitemShut {NoStop}%
\bibitem [{\citenamefont {Mahan}(2013)}]{Mahan2013}%
  \BibitemOpen
  \bibfield  {author} {\bibinfo {author} {\bibfnamefont {G.}~\bibnamefont
  {Mahan}},\ }\href {https://books.google.de/books?id=TFDUBwAAQBAJ} {\emph
  {\bibinfo {title} {Many-Particle Physics}}},\ Physics of Solids and Liquids\
  (\bibinfo  {publisher} {Springer US},\ \bibinfo {year} {2013})\BibitemShut
  {NoStop}%
\bibitem [{\citenamefont {Chaikin}\ and\ \citenamefont
  {Lubensky}(2000)}]{Chaikin2000}%
  \BibitemOpen
  \bibfield  {author} {\bibinfo {author} {\bibfnamefont {P.~M.}\ \bibnamefont
  {Chaikin}}\ and\ \bibinfo {author} {\bibfnamefont {T.~C.}\ \bibnamefont
  {Lubensky}},\ }\href {https://books.google.de/books?id=P9YjNjzr9OIC} {\emph
  {\bibinfo {title} {{Principles of Condensed Matter Physics}}}}\ (\bibinfo
  {publisher} {Cambridge University Press},\ \bibinfo {year}
  {2000})\BibitemShut {NoStop}%
\bibitem [{\citenamefont {Ma}(1973)}]{Ma1973}%
  \BibitemOpen
  \bibfield  {author} {\bibinfo {author} {\bibfnamefont {S.-k.}\ \bibnamefont
  {Ma}},\ }\href {\doibase 10.1103/PhysRevA.7.2172} {\bibfield  {journal}
  {\bibinfo  {journal} {Phys. Rev. A}\ }\textbf {\bibinfo {volume} {7}},\
  \bibinfo {pages} {2172} (\bibinfo {year} {1973})}\BibitemShut {NoStop}%
\bibitem [{\citenamefont {Coleman}(1988)}]{Coleman1988}%
  \BibitemOpen
  \bibfield  {author} {\bibinfo {author} {\bibfnamefont {S.}~\bibnamefont
  {Coleman}},\ }\href {https://books.google.de/books?id=iLwgAwAAQBAJ} {\emph
  {\bibinfo {title} {Aspects of Symmetry: Selected Erice Lectures}}}\ (\bibinfo
   {publisher} {Cambridge University Press},\ \bibinfo {year}
  {1988})\BibitemShut {NoStop}%
\bibitem [{\citenamefont {Zinn-Justin}(2002)}]{ZinnJustin2002}%
  \BibitemOpen
  \bibfield  {author} {\bibinfo {author} {\bibfnamefont {J.}~\bibnamefont
  {Zinn-Justin}},\ }\href {https://books.google.de/books?id=N8DBpTzBCJYC}
  {\emph {\bibinfo {title} {Quantum Field Theory and Critical Phenomena}}},\
  International series of monographs on physics\ (\bibinfo  {publisher}
  {Clarendon Press},\ \bibinfo {year} {2002})\BibitemShut {NoStop}%
\bibitem [{Note3()}]{Note3}%
  \BibitemOpen
  \bibinfo {note} {We note that Eq.~\protect \textup {\hbox {\mathsurround \z@
  \protect \normalfont (\ignorespaces \ref {eq:rho_crit}\unskip \@@italiccorr
  )}} is qualitatively different from the corresponding expression obtained for
  the 3D-XY transition. In the present case, the complex order parameter is
  given by $\sigma +i w$, where only $\sigma $ undergoes a phase transition due
  to the finite gap $r_w$. This difference manifests itself in the fact that
  for finite $r_w$, $\rho _s^{(0)}$ approaches a constant for $k_z\to 0$
  (corresponding to a superconducting response). On the other hand, for the
  XY-transition (with $r_w=0$) the leading term is proportional to $k_z$
  (corresponding to a dissipative response).}\BibitemShut {Stop}%
\bibitem [{Note4()}]{Note4}%
  \BibitemOpen
  \bibinfo {note} {At some point in the parameter space, $\protect \mathaccentV
  {bar}016{n}_s$ saturates, but it is outside the validity of our field theory
  analysis.}\BibitemShut {Stop}%
\bibitem [{\citenamefont {Mourik}\ \emph {et~al.}(2012)\citenamefont {Mourik},
  \citenamefont {Zuo}, \citenamefont {Frolov}, \citenamefont {Plissard},
  \citenamefont {Bakkers},\ and\ \citenamefont {Kouwenhoven}}]{Mourik2012}%
  \BibitemOpen
  \bibfield  {author} {\bibinfo {author} {\bibfnamefont {V.}~\bibnamefont
  {Mourik}}, \bibinfo {author} {\bibfnamefont {K.}~\bibnamefont {Zuo}},
  \bibinfo {author} {\bibfnamefont {S.~M.}\ \bibnamefont {Frolov}}, \bibinfo
  {author} {\bibfnamefont {S.~R.}\ \bibnamefont {Plissard}}, \bibinfo {author}
  {\bibfnamefont {E.~P. a.~M.}\ \bibnamefont {Bakkers}}, \ and\ \bibinfo
  {author} {\bibfnamefont {L.~P.}\ \bibnamefont {Kouwenhoven}},\ }\href
  {\doibase 10.1126/science.1222360} {\bibfield  {journal} {\bibinfo  {journal}
  {Science}\ }\textbf {\bibinfo {volume} {336}},\ \bibinfo {pages} {1003}
  (\bibinfo {year} {2012})}\BibitemShut {NoStop}%
\bibitem [{\citenamefont {Albrecht}\ \emph {et~al.}(2016)\citenamefont
  {Albrecht}, \citenamefont {Higginbotham}, \citenamefont {Madsen},
  \citenamefont {Kuemmeth}, \citenamefont {Jespersen}, \citenamefont {Nyg},
  \citenamefont {Krogstrup},\ and\ \citenamefont {Marcus}}]{Albrecht2016}%
  \BibitemOpen
  \bibfield  {author} {\bibinfo {author} {\bibfnamefont {S.~M.}\ \bibnamefont
  {Albrecht}}, \bibinfo {author} {\bibfnamefont {A.~P.}\ \bibnamefont
  {Higginbotham}}, \bibinfo {author} {\bibfnamefont {M.}~\bibnamefont
  {Madsen}}, \bibinfo {author} {\bibfnamefont {F.}~\bibnamefont {Kuemmeth}},
  \bibinfo {author} {\bibfnamefont {T.~S.}\ \bibnamefont {Jespersen}}, \bibinfo
  {author} {\bibfnamefont {J.}~\bibnamefont {Nyg}}, \bibinfo {author}
  {\bibfnamefont {P.}~\bibnamefont {Krogstrup}}, \ and\ \bibinfo {author}
  {\bibfnamefont {C.~M.}\ \bibnamefont {Marcus}},\ }\href {\doibase
  10.1038/nature17162} {\bibfield  {journal} {\bibinfo  {journal} {Nature}\
  }\textbf {\bibinfo {volume} {531}},\ \bibinfo {pages} {206} (\bibinfo {year}
  {2016})}\BibitemShut {NoStop}%
\end{thebibliography}%
\end{document}